\documentstyle[preprint,prb,aps]{revtex}

\newcommand{\vnu}{V_{{\bf 0}}}
\newcommand{\nnu}{n_{{\bf 0}}}
\newcommand{\hsi}{\hbar\Sigma_{12}}

\begin{document}
\draft
\title{Variational approach to the dilute Bose gas}
\author{M. Bijlsma and H.T.C. Stoof}
\address{University of Utrecht, Institute for Theoretical
Physics, Princetonplein 5, P.O. Box 80.006\\ 3508 TA Utrecht, The Netherlands\\
}

\maketitle
\begin{abstract}
We study the weakly-interacting Bose gas in both two and three dimensions using
a variational approach. In particular we construct the thermodynamic potential
of the gas to within ladder approximation and find by minimization an accurate
mean-field description of the dilute Bose gas. Using spin-polarized atomic
hydrogen as a specific example, we
obtain an improved description of the Bose-Einstein condensed phase in three
dimensions and a signature of a phase transition in two dimensions. The latter
cannot be found by straightforward application of perturbation theory around
the ideal Bose gas.
\end{abstract}
\vspace{3cm}
\pacs{PACS numbers: 64.10.+h, 67.65.+z, 67.40.-w}
%\newpage

\section{Introduction}
The theoretical study of a three-dimensional dilute Bose gas in the regime
where quantum degeneracy effects play a role has a long history,\cite{NNBO} and
was in first instance important for a microscopic understanding of the
$\lambda-$transition in liquid $^4He$, although quantative predictions for this
strongly-interacting system cannot be obtained by these means. The latter is
however not true for magnetically trapped quantum gases,\cite{GREK} which
explains why the
phenomenon of Bose-Einstein condensation, accompanied by the occurrence of
superfluidity, is still an active field of research today. Moreover, it turns
out that there is no completely
satisfying theory for these systems,\cite{GRIF1,GRIF2} since existing
treatments, which are
based on a Bogoliubov
kind of approach, fail for instance to show the correct order of the phase
transition and
yield a first-order transition to the condensed phase \cite{LEEY,STRAL} instead
of the second-order transition expected from the theory of critical phenomena.
In this paper we will develop a variational approach and show for example that
only a slight change in the Bogoliubov theory can resolve this issue.

The two-dimensional dilute Bose gas has become subject of study only more
recently
when it was realized that this system is in the same universality class as
the XY-model \cite{REF9} and should therefore exhibit a Kosterlitz-Thouless
phase transition.\cite{KOSTH} Physically this implies that there is no
macroscopic occupation of one single quantum
state, but there is nevertheless a transition to a superfluid state through the
mechanism of
binding of vortices with opposite vorticity. This mechanism causes the
superfluid density $n_s$ to show a universal discontinuity such that
$n_s\Lambda^2=4$ at the critical temperature,\cite{KOSN} where
$\Lambda=(2\pi\hbar^2/mk_BT)^{1/2}$ is the thermal de Broglie
wavelength and $m$ is the mass of the particles in the system. The study of
the two-dimensional Bose gas starting from a microscopic Hamiltonian was
initiated by Popov.\cite{POP1} In his approach,
which corresponds to a one-loop calculation and is based on the introduction of
a so-called quasicondensate, the superfluid density exhibits no jump, but
instead goes to zero in a continuous fashion. In addition, it has been shown
\cite{FISH,USKT} that the range of validity of Popovs approach is by no means
attainable in the experimental set-ups envisaged at present. Therefore, we have
recently extended
this work using the ladder or $T-$matrix approximation which should be
sufficiently accurate at low densities since it includes all two-body
processes.\cite{USKT} In this manner we correctly found a jump in the
superfluid density, but unfortunately, due to infrared divergences, also that
the magnitude of the quasicondensate had to be adjusted such that the jump
attained the universal value
predicted by Nelson and Kosterlitz using renormalization-group
methods.\cite{KOSN} This latter aspect of our work shows that the ladder
approximation is breaking down close to the critical temperature and needs
improvement in the critical region. This problem will be addressed in a future
publication. In this paper we will
apply our variational approach developed for the three-dimensional case also to
the two-dimensional Bose gas, because this approach can in principle also
resolve the infrared divergences in a manner which is explained in more detail
below.

The best-known experimental realization of a dilute Bose gas in which one
can try to observe either Bose-Einstein condensation or the Kosterlitz-Thouless
transition is spin-polarized atomic hydrogen. Although there is
no fundamental reason for Bose-Einstein condensation not to be attainable
in this system, the actual achievement of the critical conditions is an
extremely difficult task due to the inherent instability of the atomic
hydrogen gas against spin-flip processes which lead to particle loss and in
general also cause heating
of the gas sample. The two-dimensional configuration of course also suffers
from these drawbacks, but the critical conditions are more favorable than for
the three-dimensional case and as of the mid 1980's several experimental groups
are therefore pursuing the observation of the Kosterlitz-Thouless
transition.\cite{IKES1,SVISA,TURKU,JAPKT} In these experiments one uses liquid
$^4He$ films to confine the hydrogen atoms in one spatial direction, hence
realizing an effectively two dimensional Bose gas.

Other experimental realizations of a dilute Bose gas which are promising
candidates for the observation of quantum degeneracy effects are laser-cooled
alkali atoms and excitons. Since evaporative cooling was recently shown to work
for magnetically trapped alkalis like $Rb$ and $Na$,\cite{CORN,KETT} advances
towards the observation of Bose-Einstein condensation are particularly fast
here and recently the actual observation of a Bose-Einstein condensate in an
ultracold $^{87}Rb$ gas has been reported.\cite{SCIENCE} Recent developments in
MBE-technologies have also led to new opportunities to reach the critical
conditions for the Kosterlitz-Thouless transition in a two-dimensional
(dumbbell) exciton gas confined by a double quantum-well structure.\cite{BUTOV}
The three-dimensional exciton gas in e.g. pure $Cu_2O$ crystals has a somewhat
longer history.\cite{EXCI1} It has
even been claimed that the condensation transition has already been observed in
such a system.\cite{EXCI2} However, mainly due to the lack of a convincing
interpretation of the obtained experimental data, there is no decisive proof of
this fact as yet.

As mentioned before we will here first study the dilute Bose gas in general by
making use of a variational approach, and then take spin-polarized atomic
hydrogen as a specific example. A variational calculation is a conceptually
simple
tool \cite{SCHIF,ROSE} that has been applied to such various subjects as for
instance the investigation of the
triviality of $\lambda\phi^4$ field theories in different spacetime
dimensions,\cite{STEV1} investigation of the phase diagram of the
$O(N)$ vector model,\cite{BAMO} and construction of the ground state of the
Sine-Gordon model.\cite{ZAZOX} The
basic idea is to make a Gaussian ansatz for the ground-state wavefunctional and
its excitations, which amounts to making an ansatz
for the dispersion of the quasiparticles in the system, and then to calculate
the approximate thermodynamic potential $\Omega=\Omega_{tr}+\langle {\cal
H}-{\cal H}_{tr}\rangle_{tr}$, where ${\cal H}_{tr}$ is a trial Hamiltonian
that has the trial wavefunctionals as eigenstates,
$\Omega_{tr}=-k_BT~ln(Tr[e^{-\beta({\cal H}_{tr}-\mu{\cal N})}])$ and
$\langle...\rangle_{tr}$ is the corresponding grand canonical thermal average.
One subsequently minimizes this thermodynamic potential
with respect to the variational parameters in the Gaussian wavefunctionals, and
with respect to the expectation value of the field of interest.

In the case of the dilute Bose gas the results obtained in this manner have to
be used in the equation of state which makes for example the determination of
the critical
temperature possible. This procedure
is a mean-field approach in the sense that no fluctuations in the expectation
value of the field are taken into
account. However, it is nonperturbative, and clearly goes beyond the one-loop
level. The main advantage of using a variational approach in two dimensions is
that the notorious infrared divergences are automatically circumvented because
the dispersion relation of the particles will in general display a gap. This
gap is analogous to the (arbitrary) infrared cutoff $k_0$ that Popov introduced
to define the quasicondensate $n_0(k_0)$. Note however that in the variational
procedure minimization of the thermodynamic potential will fix the magnitude of
the gap. Hence, there will be no adjustable parameter left in the calculation,
as was the case in our previous treatment of the two-dimensional dilute Bose
gas where we had to use the universal jump relation to determine the infrared
cutoff $k_0$.\cite{USKT} This convenient feature strongly motivated us to study
also the two-dimensional system within the framework of a variational
calculation.

Our Gaussian ansatz will introduce a condensate also in two dimensions. Of
course, the existence of a true condensate in two dimensions is principally
forbidden by the Mermin-Wagner-Hohenberg theorem.\cite{MEWA} However, using
trial wavefunctionals with a broken symmetry in a variational approach can
still be usefull, because even if the form of the trial wavefunctionals
deviates considerably from the exact wavefunctionals describing the phase below
the critical temperature, the estimate for the total energy of the system can
be extremely good.

The organisation of the article is as follows. In section II we derive the
desired expression for the thermodynamic potential $\Omega$ and we show in
particular that minimization with respect to the variational parameters
automatically introduces the many-body $T-$matrix into the problem. However, it
turns out that in its most simple form the variational principle does not yield
a completely consistent picture. To resolve this problem we have to improve all
expressions to the $T-$matrix level. After this improvement we are ready to
apply the theory to atomic hydrogen. The results are presented in section III.
In section IIIA we consider the three-dimensional system and section IIIB is
devoted to
the two-dimensional system. Our approach indicates the onset of a
new phase in the latter case, which cannot be found by an application of
perturbation theory around the ideal Bose gas. We are however not able to
describe the properties of this phase by our variational method but argue that
it corresponds to the superfluid bound-vortex state. In section IV we draw some
conclusions from our work.

\section{Thermodynamic potential}
We study the dilute Bose gas using the functional integral
formulation of the grand canonical partition function.\cite{NEGO} Thus
\begin{equation}
Z_{gr}\equiv e^{-\beta\Omega}=\int
d[\psi^*]d[\psi]\exp\{-\frac{1}{\hbar}S[\psi^*,\psi]\}~,
\end{equation}
where the functional integral is over $c$-number fields
$\psi^*({\bf x},\tau)$ and $\psi({\bf x},\tau)$ periodic in
imaginary time over $\hbar\beta=\hbar/k_BT$.
The Euclidian action for the dilute Bose gas is given by
\begin{eqnarray}
S[\psi^*,\psi]&=&\int_{0}^{\hbar\beta}d\tau \left( \int d{\bf x}~
\psi^*({\bf x},\tau)\left[ \hbar\frac{\partial}{\partial\tau}-
\frac{\hbar^2{\bf \nabla}^2}{2m}-\mu \right] \psi({\bf x},\tau)\right.
\nonumber \\ &+&\left.\frac{1}{2}\int d{\bf x}\int d{\bf x'}~\psi^*({\bf
x},\tau)\psi^*({\bf
x'},\tau)V({\bf x}-{\bf x'})\psi({\bf x'},\tau)\psi({\bf x},\tau) \right)~,
\end{eqnarray}
with $\mu$ the chemical potential and $V({\bf x}-{\bf x'})$ the
effectively repulsive interaction potential, meaning that the scattering length
$a$ is positive.

Expanding the fields in Fourier modes through
\begin{equation}
\psi({\bf x},\tau)=\frac{1}{(\hbar\beta V)^{1/2}}\sum_{{\bf k},n} a_{{\bf
k},n}e^{i({\bf k\cdot x}-\omega_n\tau)}
\end{equation}
and the complex conjugate expression for $\psi^*({\bf x},\tau)$, we can
write the action in momentum space as
\begin{eqnarray}
S[a^*,a]=\sum_{{\bf k},n}(-i\hbar\omega_n+\epsilon_{{\bf k}}-\mu)
a^*_{{\bf k},n}a_{{\bf k},n}+\frac{1}{2}\frac{1}{\hbar\beta V}
\sum_{\stackrel{\bf
k,k',q}{n,n',m}} V_{\bf q}a^*_{{\bf k+q},n+m}a^*_{{\bf k'-q},n'-m}
a_{{\bf k'},n'}a_{{\bf k},n}~.
\end{eqnarray}
In this equation $\epsilon_{{\bf k}}=
\hbar^2{\bf k}^2/2m$ and $V_{\bf q}=\int d{\bf x}~V({\bf x})e^{-i{\bf q\cdot
x}}$ is the Fourier transform of the
interaction potential. Since we will be interested in the temperature
regime where the spatial extension $\Lambda$ of the hydrogen atoms
is much larger than the range of the interaction potential
($a/\Lambda\ll 1$) there is essentially only $s$-wave scattering taking place
in the gas and we
can neglect the momentum dependence of the interaction potential. Therefore we
set $V_{\bf q}=\vnu$. The momenta are restricted to either two or three
dimensions
and $V$ is the area or volume of the corresponding system. The bosonic
Matsubara frequencies $\omega_n=2\pi n/\hbar\beta$ reflect the
periodicity of the fields in (imaginary) time.

We now have the basic ingredients and can start the procedure of
variationally calculating the thermodynamic potential $\Omega$. Breaking
the symmetry by introducing $\psi({\bf x},\tau)=\sqrt{\nnu}+
\psi'({\bf x},\tau)$ and expanding the resulting expression, the action up to
quadratic order in first instance becomes
\begin{eqnarray}
S^{(2)}[a,a^*]&=&-\hbar\beta\mu\nnu V+\frac{1}{2}\hbar\beta\nnu^2\vnu V+
{\sum_{{\bf k},n}}^{'}(-i\hbar\omega_n+\epsilon_{{\bf
k}}-\mu+2\nnu\vnu)a^*_{{\bf k},n}a_{{\bf k},n}\nonumber \\
&+&\frac{1}{2}\nnu\vnu{\sum_{{\bf k},n}}^{'}(a^*_{{\bf k},n}a^*_{{\bf -k},-n}
+a_{{\bf k},n}a_{{\bf -k},-n})~,
\end{eqnarray}
where the prime denotes that ${\bf k}={\bf 0}$ is excluded from the sum.
{}From this we see that the lowest order approximations to the so-called normal
and anomalous selfenergies are respectively $\hbar\Sigma_{11}({\bf
k},\omega_n)=2\nnu\vnu$ and $\hsi({\bf k},\omega_n)=\nnu\vnu$. This gives a
strong indication to the form of the variational wavefunctional we have to take
to correctly describe the dilute Bose gas. As the choice of Gaussian
wavefunctionals is formally identical to an ansatz for the quadratic trial
Hamiltonian ${\cal H}_{tr}$ which they are supposed to diagonalize, we now
write the full action as
\begin{eqnarray}
S[a,a^*]&=&-\hbar\beta\mu\nnu V+\frac{1}{2}\hbar\beta\nnu^2\vnu V+
{\sum_{{\bf k},n}}^{'}(-i\hbar\omega_n+\epsilon_{{\bf
k}}-\mu+\hbar\Sigma_{11}({\bf k},\omega_n))a^*_{{\bf k},n}a_{{\bf
k},n}\nonumber \\
&+&\frac{1}{2}\hsi({\bf k},\omega_n){\sum_{{\bf k},n}}^{'}(a^*_{{\bf
k},n}a^*_{{\bf -k},-n}+a_{{\bf k},n}a_{{\bf -k},-n})\nonumber \\
&+&(2\nnu\vnu-\hbar\Sigma_{11}({\bf k},\omega_n)){\sum_{{\bf k},n}}^{'}
a^*_{{\bf k},n}a_{{\bf k},n}+\frac{1}{2}(\nnu\vnu-\hsi({\bf
k},\omega_n)){\sum_{{\bf k},n}}^{'}(a^*_{{\bf k},n}a^*_{{\bf -k},-n}
+a_{{\bf k},n}a_{{\bf -k},-n})\nonumber \\
&+&\sqrt{\frac{\nnu}{\hbar\beta V}}{\sum_{\stackrel{\bf k',q}{n',m}}}^{'}\vnu
a^*_{{\bf q},m}a^*_{{\bf k'-q},n'-m}a_{{\bf k'},n'}+
\sqrt{\frac{\nnu}{\hbar\beta V}}{\sum_{\stackrel{\bf k,q}{n,m}}}^{'}\vnu
a^*_{{\bf k+q},n+m}a_{{\bf q},m}a_{{\bf k},n}\nonumber \\
&+&\frac{1}{2}\frac{1}{\hbar\beta V}{\sum_{\stackrel{\bf
k,k',q}{n,n',m}}}^{'}\vnu a^*_{{\bf k+q},n+m}a^*_{{\bf k'-q},n'-m}
a_{{\bf k'},n'}a_{{\bf k},n}~,
\end{eqnarray}
where we introduced nontrivial normal and anomalous selfenergies in the first
two quadratic terms in the right-hand side of Eq.\ (6) by simply adding and
subtracting $\sum'_{{\bf k},n}(\hbar\Sigma_{11}({\bf k},\omega_n)a^*_{{\bf
k},n}a_{{\bf k},n}+\frac{1}{2}\hsi ({\bf k},\omega_n)(a^*_{{\bf k},n}a^*_{{\bf
-k},-n}+a_{{\bf k},n}a_{{\bf -k},-n}))$. The quantities $\hbar\Sigma_{11}({\bf
k},\omega_n)$ and $\hsi ({\bf k},\omega_n)$ are now considered as the
variational parameters of the trial wavefunctionals and are determined from a
minimization of the thermodynamic potential. As indicated by their arguments,
the selfenergies are in principle functions of momentum and frequency. However,
since we are dealing with a dilute gas in the limit where only $s$-wave
scattering is of importance, we will neglect these dependencies and henceforth
omit the arguments. Note furthermore that we assume the selfenergies to be
real, hence $\hsi=\hbar\Sigma_{21}$, which amounts to a choice of gauge that is
related to our choice of $\langle\psi\rangle=\sqrt{\nnu}$ being real.

Performing the Bogoliubov transformation to diagonalize the trial action
\begin{eqnarray}
S^{(tr)}[a^*,a]&=&{\sum_{{\bf k},n}}^{'}(-i\hbar\omega_n+\epsilon_{{\bf
k}}-\mu+\hbar\Sigma_{11})a^*_{{\bf k},n}a_{{\bf
k},n}+\frac{1}{2}\hsi{\sum_{{\bf k},n}}^{'}(a^*_{{\bf k},n}a^*_{{\bf
-k},-n}+a_{{\bf k},n}a_{{\bf -k},-n})~,
\end{eqnarray}
we find that the dispersion relation for the Bogoliubov quasiparticles is given
by
\begin{equation}
\hbar\Omega_{{\bf k}}=\sqrt{(\epsilon_{{\bf k}}-\mu')^2-(\hsi)^2}~,
\end{equation}
where we have introduced $\mu'=\mu-\hbar\Sigma_{11}$. Moreover, the usual
coherence factors \cite{FETWAL} for this transformation are given by $u_{\bf
k}=\frac{1}{2}[\sqrt{(\epsilon_{\bf k} -\mu'+\hsi)/\hbar\Omega_{\bf
k}}+\sqrt{\hbar\Omega_{\bf k}/(\epsilon_{\bf k}-\mu'+\hsi)}]$ and $v_{\bf
k}=\frac{1}{2}[\sqrt{(\epsilon_{\bf k}-\mu'+\hsi)/\hbar\Omega_{\bf
k}}-\sqrt{\hbar\Omega_{\bf k}/(\epsilon_{\bf k}-\mu'+\hsi)}]$ respectively. The
Hugenholtz-Pines theorem \cite{HUPI} states that the exact normal and anomalous
selfenergies satisfy
\begin{equation}
\mu=\hbar\Sigma_{11}({\bf 0},0)-\hsi({\bf 0},0)~,
\end{equation}
leading to a gapless dispersion. This condition is however $not$ imposed in our
variational calculation since the Hugenholtz-Pines relation is only valid in a
stationary point of the action \cite{USKT} and the variational approach by
definition also describes nonstationary points. As we will see below, Eq.\ (9)
will nonetheless turn out to hold for the three-dimensional case if the
temperatures are not too close to absolute zero. In two dimensions Eq.\ (9)
will not be valid for temperatures unequal to zero since it would lead to
infrared diverging integrals in the thermodynamic potential and is therefore
excluded from the solution space by the variational principle. As mentioned
above, this was actually our main motivation for applying this approach also to
the two-dimensional Bose gas.

The next step now is to calculate from the functional integral the
thermodynamic potential $\Omega$ which equals
\begin{equation}
\Omega=-\frac{1}{\beta}ln Z_{tr} +\frac{1}{\hbar\beta}\langle
S^{(int)}\rangle_{tr}~,
\end{equation}
where $S^{(int)}$ is the interaction term $S-S^{(tr)}$ found by a combination
of Eq.\ (6) and Eq.\ (7).
Using again the above mentioned Bogoliubov transformation a straightforward
calculation of this quantity leads to
\begin{eqnarray}
\frac{1}{V}\Omega(\mu',\hsi,\nnu;\mu,T)&=&\frac{1}{2}\nnu^2\vnu-\mu
n+\frac{k_BT}{V}{\sum_{{\bf k}}}^{'}ln(1-e^{-\beta\hbar\Omega_{{\bf
k}}})+\frac{1}{2}\frac{1}{V}{\sum_{{\bf k}}}^{'}(\hbar\Omega_{{\bf
k}}-\epsilon_{{\bf k}}+\mu')\nonumber \\ &+&(2\nnu\vnu+\mu')
(n-\nnu)+(\hsi-\nnu\vnu)\hsi\alpha\nonumber \\
&+&\vnu(n-\nnu)^2+\frac{1}{2}\frac{1}{V}\vnu(\hsi)^2\alpha^2~,
\end{eqnarray}
where
\begin{eqnarray}
n=\nnu+\frac{1}{V}{\sum_{{\bf k},n}}^{'}\langle a^*_{{\bf k},n}a_{{\bf
k},n}\rangle=\nnu+\frac{1}{V}{\sum_{\bf k}}^{'}(\frac{\epsilon_{\bf
k}-\mu'}{\hbar \Omega_{\bf k}}N(\hbar\Omega_{\bf k})+\frac{\epsilon_{\bf
k}-\mu'-\hbar\Omega_{\bf k}}{2\hbar \Omega_{\bf k}})
\end{eqnarray}
is the equation for the density and the dispersion $\hbar\Omega_{\bf k}$ is
given by Eq.\ (8). Furthermore, $\alpha$ is introduced as a shorthand notation
for
\begin{eqnarray}
\frac{1}{\hsi}\frac{1}{V}{\sum_{{\bf k},n}}^{'}\langle a^*_{{\bf k},n}a^*_{{\bf
-k},-n}\rangle=\frac{1}{V}{\sum_{\bf k}}^{'}\frac{1+2N(\hbar\Omega_{\bf
k})}{2\hbar\Omega_{\bf k}}~,
\end{eqnarray}
and $N(x)=1/(e^{\beta x}-1)$ is the Bose-Einstein distribution function.

We now  first of all have to minimize this expression for fixed $\nnu$,
chemical potential $\mu$ and temperature $T$ with respect to the variational
parameters $\hsi$ and $\hbar\Sigma_{11}$, or equivalently $\hsi$ and $\mu'$.
Putting $(\partial\Omega/\partial\hsi)=0$ and $(\partial\Omega/\partial\mu')=0$
yields respectively
\begin{equation}
\hsi=\nnu\frac{\vnu}{1+\vnu\alpha}-\frac{2n\vnu+\mu'-\mu}{(1+\vnu\alpha)(\alpha+\hsi\frac{\partial\alpha}{\partial\hsi})}\frac{\partial n}{\partial\hsi}
\end{equation}
and
\begin{equation}
(2n\vnu+\mu'-\mu)\frac{\partial
n}{\partial\mu'}=\left[\hsi(1+\vnu\alpha)-\nnu\vnu\right]\hsi
\frac{\partial\alpha}{\partial\mu'}~,
\end{equation}
which are both solved by
\begin{equation}
\hsi=\nnu\frac{\vnu}{1+\vnu\alpha}
\end{equation}
and
\begin{equation}
\mu'=\mu-2n\vnu~.
\end{equation}
Calculation of the second derivatives of the thermodynamic potential indeed
shows that this solution corresponds to a minimum for any $\vnu>0$. For
$\vnu<0$ the obtained solution would correspond to a maximum in the
thermodynamic potential, implying an unstable solution. This is physically
reasonable since a negative $\vnu$ implies an effective attraction between the
constituent particles and thus a negative scattering length $a$. As a result, a
magnetically trapped gas indeed turns out to be unstable against the formation
of a dense (solid or liquid) phase in the (local density) limit that the
temperatures are much larger than the splittings $\hbar\omega$ of the
one-particle eigenstates in the trap, and the correlation length is smaller
that the typical trap size.\cite{HENE} Quantatively this means that
$k_BT/\hbar\omega\gg l/\mid a\mid\gg 1$, where $l=\sqrt{\hbar/m\omega}$ is the
extent of the lowest lying unperturbed harmonic oscillator state of the trap.
However, in the opposite limit it has recently been shown by a numerical
solution of the nonlinear Schr\"{o}dinger equation that a condensate may be
metastable for a sufficiently small occupation number.\cite{KEIT1} Although
this is a very interesting result in the light of the fact that a number of
alkali atoms, e.g.\ $Cs$ and $Li$, have negative scattering
lengths,\cite{DNARD1,DNARD2} we will not consider this highly inhomogeneous
case here and therefore restrict ourselves to $a>0$.

Starting with the second equation we immediately conclude that the solution for
the normal selfenergy is $\hbar\Sigma_{11}=2n\vnu$, meaning that the
variational calculation yields for this particular variable only the one-loop
(i.e.\ Hartree-Fock) expression. The first equation shows however, that the
anomalous selfenergy $\hsi$ is obtained in an approximation that goes far
beyond one loop. In fact, as we will now show, does it constitute the many-body
$T-$matrix approximation to the selfenergy $\hsi$.\cite{USKT}

The many-body $T-$matrix $T^{MB}({\bf k},{\bf k'},{\bf K};z)$, which is
diagrammatically depicted in Fig.\ 1(a), describes the collision of two atoms
having relative momenta ${\bf k'}$ and ${\bf k}$ before, respectively after the
collision, and a total center-of-mass momentum ${\bf K}$ and center-of-mass
energy $z$ in an environment containing the same atoms. The equation for the
many-body $T-$matrix, which by construction sums all the ladder diagrams, can
easily be derived if one realizes that the free propagation of the atoms is
determined by the trial Hamiltonian, that can be diagonalized by a Bogoliubov
transformation. A somewhat lengthy but straightforward calculation gives in the
thermodynamic limit
\begin{eqnarray}
T^{MB}({\bf k},{\bf k'},{\bf K};z)&=&V({\bf k}-{\bf k'})+\int \frac{d{\bf
k''}}{(2\pi)^d} V({\bf k}-{\bf k''})\nonumber \\
&\times&\left\{\left[\frac{u_+^2u_-^2}{z-\hbar\Omega_+-\hbar\Omega_-}-
\frac{v_+^2v_-^2}{z+\hbar\Omega_++\hbar\Omega_-}\right]\right.(1+N_++N_-)\nonumber \\ &+&\left[\left.\frac{u_-^2v_+^2}{z+\hbar\Omega_+-\hbar\Omega_-}-\frac{u_+^2v_-^2}{z-\hbar\Omega_++\hbar\Omega_-}\right](N_+-N_-)\right\}\nonumber \\&\times&T^{MB}({\bf k''},{\bf k'},{\bf K};z)~,
\end{eqnarray}
where $N_+\equiv N(\hbar\Omega_+)$ and $N_-\equiv N(\hbar\Omega_-)$. The plus
sign denotes the momentum argument $\frac{{\bf K}}{2}+{\bf k''}$, and similarly
the minus sign denotes the argument $\frac{{\bf K}}{2}-{\bf k''}$. This
expression for the many-body $T-$matrix has a clear physical interpretation:
Two incoming atoms with momenta $\frac{{\bf K}}{2}+{\bf k''}$ and $\frac{{\bf
K}}{2}-{\bf k''}$ can scatter out of their momentum states either by the
creation of two quasi-particles, by the creation of two quasi-holes and by the
creation of one quasiparticle and one quasi-hole. The net rates for the first
two processes are proportional to $u_+^2u_-^2[(1+N_+)(1+N_-)-N_+N_-]$ and
$v_+^2v_-^2[N_+N_--(1+N_+)(1+N_-)]$, respectively, if we take the usual Bose
enhancement factors for the scattering into occupied states into account. For
the last process the quasi-particle can have either momentum $\frac{{\bf
K}}{2}+{\bf k''}$ or $\frac{{\bf K}}{2}-{\bf k''}$. The net rate for these
processes are proportional to $u_-^2v_+^2[(1+N_-)N_+-N_-(1+N_+)]$ and
$u_+^2v_-^2[(1+N_+)N_--N_+(1+N_-)]$, respectively. Together with the usual
energy denominators, which are well-known from second-order perturbation
theory, this fully explains the structure of the right-hand side of Eq.\ (18).

Since we neglected all momentum dependencies of the selfenergies
$\hbar\Sigma_{11}$ and $\hsi$ in the variational approach, we will also do so
in the above equation for the $T-$matrix. When we furthermore set $u_{\bf k}=1$
and $v_{\bf k}=0$ by making use of the fact that the Bogoliubov spectrum
$\hbar\Omega_{\bf k}$ deviates from $\epsilon_{\bf k}-\mu'$ only for a very
small momentum interval such that $\epsilon_{\bf
k}\stackrel{<}{\sim}\hsi<n\vnu$, we find that Eq.\ (18) reduces to
\begin{equation}
T^{MB}({\bf 0},{\bf 0},{\bf 0};z)=\vnu+\vnu T^{MB}({\bf 0},{\bf 0},{\bf
0};z)\int\frac{d{\bf k}}{(2\pi)^d}\frac{1+2N(\hbar\Omega_{\bf
k})}{z-2\hbar\Omega_{\bf k}}~,
\end{equation}
with an error of $O(\vnu na\Lambda^2)$, which is in general negligible except
in an extremely small interval of temperatures near absolute zero. By virtue of
the fact that we neglected the momentum dependence of the scattering potential
this equation can easily be solved to give
\begin{equation}
T^{MB}({\bf 0},{\bf 0},{\bf 0};0)=\frac{\vnu}{1+\vnu\alpha}~,
\end{equation}
which finally establishes our claim that the variational calculation gives the
anomalous selfenergy to within $T-$matrix approximation, i.e. $\hsi=\nnu
T^{MB}({\bf 0},{\bf 0},{\bf 0};0)$. Furthermore, Eq.\ (20) implies that
$\alpha=1/T^{MB}({\bf 0},{\bf 0},{\bf 0};0)-1/\vnu$.

Note that in the calculation we are carrying out, this $T-$matrix still depends
on the variational parameters $\hbar\Sigma_{11}$ and $\hsi$ because they enter
into the dispersion $\hbar\Omega_{\bf k}$. Moreover, the many-body $T-$matrix
can be related to the two-body $T-$matrix, which describes the collision of two
particles in a vacuum instead of a medium, through \cite{USKT}
\begin{equation}
\frac{1}{T^{MB}({\bf 0},{\bf 0},{\bf 0};0)}=\frac{1}{T^{2B}({\bf 0},{\bf
0};-2\mu')}+\frac{1}{V}\sum_{\bf k}\frac{N(\hbar\Omega_{\bf
k})}{\hbar\Omega_{\bf k}}~.
\end{equation}
The low-energy limit ($\epsilon_{\bf k}\ll\hbar^2/2ma^2$) of the two-body
$T-$matrix has been studied extensively in the literature.\cite{STOG} When
configuration space is three-dimensional it is accurately represented by
\begin{equation}
T^{2B}({\bf 0},{\bf 0};\epsilon_{\bf k})=\frac{4\pi a \hbar^2}{m}~,
\end{equation}
whereas
\begin{equation}
T^{2B}({\bf 0},{\bf 0};\epsilon_{\bf k})=\frac{4\pi\hbar^2/m}{\pi
i-ln(k^2a^2/8)-2\gamma}
\end{equation}
when configuration space is two-dimensional. In these expressions $a$ denotes
the scattering length or effective hard sphere radius of the interaction
potential and $\gamma\simeq 0.5772$ is Euler's constant. These long wavelength
expressions will enable us in a later stage to formulate our final results
solely in terms of the scattering length instead of the full (microscopic)
potential $V({\bf x})$.

Comparing Eqs.\ (16) and (17) we conclude that the variational calculation as
performed above leads to an inconsistent theory: The normal selfenergy is
obtained to one-loop order only, whereas the anomalous selfenergy is obtained
in the $T-$matrix approximation, which contains any number of loops. The
inconsistency can be better understood by diagrammatically analyzing the
structure of Eqs.\ (16) and (17), making use of Eq.\ (12) for the density and
the definition of $\alpha$ given in Eq.\ (13). The result is depicted in Fig.\
1(b) together with the Dyson equation for the normal and anomalous Greens
functions. It is clear from these diagrammatic equations that iteration will
generate the ladder series for the anomalous selfenergy $\hsi$, but not for the
normal selfenergy $\hbar\Sigma_{11}$. Moreover, it shows that the set of
diagrams which is being summed in our case of a complex $\mid\psi\mid^4$ theory
gives at most the $T-$matrix. This is in contrast to for example the result
obtained by Barnes and Ghandour.\cite{BARG} They show that a similar
variational calculation for a real scalar $\phi^4$ theory yields selfenergies
containing the ladder as well as the bubble diagrams.

To lift the inconsistency and to obtain an expression for the thermodynamic
potential independent of the (microscopic) variable $\vnu$ we have to improve
the derived expressions. We proceed as follows. Since it was shown that for any
$\vnu>0$ the solution to the minimization procedure yields a minimum in the
surface of the thermodynamic potential, we can replace it everywhere, except in
the condensate contribution $\frac{1}{2}\nnu^2\vnu$ for reasons that will
become clear shortly, by some constant $A$. Performing again the minimization
procedure now leads to
$\hsi=\nnu A/(1+A\alpha)$ and $\mu'=\mu-2nA$, where $\alpha=1/T^{MB}({\bf
0},{\bf 0},{\bf 0};0)-1/A$. It is readily seen that taking $A$ equal to the
many-body $T-$matrix evaluated at the minimum of $\Omega$ located at the point
in the $\mu'-\hsi$ plane given by the above solutions, indeed leads to the
desired selfenergies
\begin{eqnarray}
\hsi=\nnu T^{MB}({\bf 0},{\bf 0},{\bf 0};0)~,\\
\hbar\Sigma_{11}=2nT^{MB}({\bf 0},{\bf 0},{\bf 0};0)
\end{eqnarray}
in $T-$matrix approximation.

The thermodynamic potential at this minimum in first instance equals
\begin{eqnarray}
\frac{1}{V}\Omega(\nnu;\mu,T)&=&\frac{1}{2}\nnu^2\vnu-\mu
n+\frac{k_BT}{V}\sum_{{\bf k}}ln(1-e^{-\beta\hbar\Omega_{{\bf k}}})\nonumber \\
&+&\frac{1}{2}\frac{1}{V}{\sum_{{\bf k}}}^{'}(\hbar\Omega_{{\bf
k}}-\epsilon_{{\bf k}}+\mu')+(2\nnu T^{MB}({\bf 0},{\bf 0},{\bf 0};0)+\mu')
(n-\nnu)\nonumber \\ &+&T^{MB}({\bf 0},{\bf 0},{\bf 0};0)(n-\nnu)^2~,
\end{eqnarray}
from which we still have to eliminate the $\vnu$ dependence occurring in the
first term, as well as the ultraviolet divergence in the fourth term of the
right-hand side. Interestingly, these two matters are intimately connected, and
can both be resolved by writing the sum of these two terms as
\begin{eqnarray}
\frac{1}{2}\nnu^2\vnu+\frac{1}{2}\frac{1}{V}{\sum_{{\bf
k}}}^{'}(\hbar\Omega_{{\bf k}}-\epsilon_{{\bf
k}}+\mu'+\frac{(\hsi)^2}{2\hbar\Omega_{\bf
k}})-\frac{1}{2}(\hsi)^2\frac{1}{V}{\sum_{\bf k}}^{'}\frac{1}{2\hbar\Omega_{\bf
k}}~. \nonumber
\end{eqnarray}
By also adding and subtracting $\frac{1}{2}(\hsi)^2\frac{1}{V}\sum_{\bf
k}\frac{N(\hbar\Omega_{\bf k})}{\hbar\Omega_{\bf k}}$ and realizing that
$\hsi=\nnu T^{MB}({\bf 0},{\bf 0},{\bf 0};0)$, we see that $\vnu$ is exactly
renormalized to the many-body $T-$matrix if one of the $T^{MB}$'s in $(\hsi)^2$
is replaced by $\vnu$, since then
\begin{equation}
\frac{1}{2}\nnu^2\vnu-\frac{1}{2}\nnu^2\vnu T^{MB}({\bf 0},{\bf 0},{\bf
0};0)\frac{1}{V}{\sum_{\bf k}}^{'}\frac{1+2N(\hbar\Omega_{\bf
k})}{2\hbar\Omega_{\bf k}}=\frac{1}{2}\nnu^2 T^{MB}({\bf 0},{\bf 0},{\bf 0};0)
\end{equation}
due to Eq.\ (19). The final expression for the thermodynamic potential thus
becomes
\begin{eqnarray}
\frac{1}{V}\Omega(\nnu;\mu,T)&=&\frac{1}{2}\nnu^2 T^{MB}({\bf 0},{\bf 0},{\bf
0};0)-\mu n+\frac{k_BT}{V}\sum_{{\bf k}}ln(1-e^{-\beta\hbar\Omega_{{\bf
k}}})\nonumber \\&+&\frac{1}{2}\frac{1}{V}{\sum_{{\bf
k}}}^{'}(\hbar\Omega_{{\bf k}}-\epsilon_{{\bf
k}}+\mu'+\frac{(\hsi)^2}{2\hbar\Omega_{\bf
k}})+\frac{1}{2}(\hsi)^2\frac{1}{V}{\sum_{\bf k}}'\frac{N(\hbar\Omega_{\bf
k})}{\hbar\Omega_{\bf k}} \nonumber \\ &+&(2\nnu T^{MB}({\bf 0},{\bf 0},{\bf
0};0)+\mu') (n-\nnu)+T^{MB}({\bf 0},{\bf 0},{\bf 0};0)(n-\nnu)^2~,
\end{eqnarray}
with $\hsi$ given by Eq.\ (24) and $\mu'=\mu-2nT^{MB}({\bf 0},{\bf 0},{\bf
0};0)$. Note that this derivation by no means depends on the validity or
violation of the Hugenholtz-Pines relation, and that we now have obtained the
thermodynamic potential containing only the many-body $T-$matrix and not the
interaction parameter $\vnu$. The fourth term on the right-hand side can be
shown to be at most of $O(\sqrt{na^3})$ in three dimensions, and of
$O(1/ln(1/na^2))$ in two dimensions, and will henceforth be neglected.

Although the above procedure might in first instance seem somewhat heuristic,
it has indeed led to the correct expression for the thermodynamic potential in
$T-$matrix approximation. This we will now show by deriving Eq.\ (28) directly
from the original action $S[a^*,a]$ of which we have in Eq.\ (5) written down
the quadratic part. To do so we treat the off-diagonal part of $S^{(2)}$ also
as a perturbation, and consider only the diagonal part as the unperturbed
action. Thus $\langle a^*_{{\bf k},n}a_{{\bf
k},n}\rangle=-\hbar/(i\hbar\omega_n-\epsilon_{\bf k}+\mu')$. In first instance
we would then immediately write down that
\begin{eqnarray}
\frac{1}{V}\Omega&=&\frac{1}{2}\nnu^2\vnu-\mu\nnu+\frac{k_BT}{V}{\sum_{{\bf
k}}}^{'}ln(1-e^{-\beta(\epsilon_{\bf k}-\mu')})-T^{MB}({\bf 0},{\bf 0},{\bf
0};0)(n-\nnu)^2~,
\end{eqnarray}
where the last term compensates for the double-counting of the interactions.
The first term can now be improved to the $T-$matrix level by performing
perturbation theory in the off-diagonal parts of the action $S^{(2)}$. The
lowest-order term contributes $-\frac{1}{2}(\hsi)^2\frac{1}{V}\sum_{\bf
k}\frac{1+2N(\epsilon_{\bf k}-\mu')}{2(\epsilon_{\bf k}-\mu')}$ to the
thermodynamic potential, whereas the consecutive terms are of $O(\nnu a^3)$
resp. $O(1/ln(1/\nnu a^2))$ and negligible. Combining this lowest-order term
with $\frac{1}{2}\nnu^2\vnu$ the thermodynamic potential now reads
\begin{eqnarray}
\frac{1}{V}\Omega&=&\frac{1}{2}\nnu^2T^{MB}({\bf 0},{\bf 0},{\bf
0};0)-\mu\nnu+\frac{k_BT}{V}{\sum_{{\bf k}}}^{'}ln(1-e^{-\beta(\epsilon_{\bf
k}-\mu')})-T^{MB}({\bf 0},{\bf 0},{\bf 0};0)(n-\nnu)^2~.
\end{eqnarray}
This result now has to be compared with Eq.\ (28). Firstly we have to show,
lest they be equal, that
\begin{eqnarray}
\frac{k_BT}{V}{\sum_{{\bf k}}}^{'}ln(1-e^{-\beta\hbar\Omega_{{\bf k}}})+
\frac{1}{2}(\hsi)^2\frac{1}{V}{\sum_{\bf k}}^{'}\frac{N(\hbar\Omega_{\bf
k})}{\hbar\Omega_{\bf k}}=\frac{k_BT}{V}{\sum_{{\bf
k}}}^{'}ln(1-e^{-\beta(\epsilon_{\bf k}-\mu')})~.
\end{eqnarray}
That this equality holds to a very good approximation can be demonstrated
simply by performing an expansion of $\hbar\Omega_{\bf k}$, which leads to
\begin{equation}
ln(1-e^{-\beta\hbar\Omega_{{\bf k}}})=ln(1-e^{-\beta(\epsilon_{\bf
k}-\mu')})-\beta\frac{(\hsi)^2}{2(\epsilon_{\bf k}-\mu')}N(\epsilon_{\bf
k}-\mu')+h.o.t.
\end{equation}
where the ratio of the higher order terms and the first order term in the
expansion is either of $O(\nnu a\Lambda^2)$ or of $O((\nnu\Lambda^2/ln(1/\nnu
a^2)))$ in the regime where these parameters are small. Close to zero
temperature these parameters become large, but what has to be compared in that
situation is the contribution of these terms to the {\em total} thermodynamic
potential. Due to the Bose factors their contribution becomes negligible.
Secondly, the two respective terms $(n-\nnu)^2$ in Eqs.\ (28) and (30) also
differ due to the difference in dispersion. Again expanding $\hbar\Omega_{\bf
k}$, one can show that the error here is of the same order as above and
negligible. Thus Eq.\ (28) obtained from the modified variational calculation
is indeed the correct result for the thermodynamic potential.

The above argument shows that the results obtained in Eqs.\ (24), (25) and (28)
can also be interpreted in a different manner. First we break the symmetry of
the theory by hand. We subsequently calculate the normal and anomalous
selfenergies $\hbar\Sigma_{11}$ and $\hsi$ in the $T-$matrix approximation.
This straightforwardly gives Eqs.\ (24) and (25) (see also Ref. 12). We then
also calculate the thermodynamic potential in $T-$matrix approximation,
yielding as we have seen Eq.\ (28). Thus we again arrive without reference to a
variational calculation at the same set of equations and
in particular at the same thermodynamic potential as a function of condensate
density, chemical potential and temperature.

The final step we now have to take is to find the value of $\nnu$ which
minimizes the thermodynamic potential at given $\mu$ and $T$. After this the
properties of the system are fully determined and it is possible to calculate
e.g.\ the condensate fraction as a function of temperature and the superfluid
density. Using the fundamental thermodynamic identity $E=\Omega+TS+\mu N$, with
the entropy $S$ given by
\begin{equation}
S=-k_B\sum_{\bf k}[N(\hbar\Omega_{\bf k})lnN(\hbar\Omega_{\bf
k})-(1+N(\hbar\Omega_{\bf k}))ln(1+N(\hbar\Omega_{\bf k}))]~,
\end{equation}
we can also calculate the specific heat from $C_V=(\frac{\partial E}{\partial
T})_N$.

Because this program has to be implemented numerically, we now restrict
ourselves to the specific example of spin-polarized hydrogen atoms in both a
three-dimensional (section IIIA) and a two-dimensional (section IIIB)
configuration. In the case of the system being three-dimensional, it leads to
improvements on existing approaches to this problem, and we therefore in
particular present results for the condensate fraction, the specific heat and
the $p-n^{-1}$ diagram.

\section{Spin-polarized atomic hydrogen}
In order to obtain Bose-Einstein condensation one magnetically traps hydrogen
atoms in the doubly polarized $\mid
d\rangle=\mid\uparrow\uparrow\hspace*{-.182in}-\rangle$ state of the upper
hyperfine manifold, resulting in a well-known effectively repulsive triplet
interaction between the atoms with a value of the scattering length of
1.34$a_0$.\cite{STOK} Typical sample densities are $10^{13}$ to $10^{14}$ atoms
per cubic centimeter with a corresponding critical temperature on the order of
tens of microkelvins.

For the two-dimensional system one uses atoms in the also doubly polarized
$\mid b\rangle =\mid\downarrow\downarrow\hspace*{-.185in}-\rangle$ state of the
lower hyperfine manifold \cite{GREK} adsorbed on a superfluid helium film. The
effective interaction between the adsorbed atoms is in a very good
approximation given by the weigthed average of the volume triplet potential,
using the square of the bound state wavefunction pertaining to the motion
perpendicular to the surface, as a weighting factor.\cite{EIJV} This procedure
leads to a two-dimensional scattering length with a value of
2.40$a_0$.\cite{STOG} The densities one aims at in this experimental setup are
typically some $10^{13}$ atoms per square centimeter, which corresponds to a
critical temperature in the range of 50 to 100 millikelvin.

\subsection{The three dimensional case}
In Fig.\ 2 we show for three different values of $\nnu a^3$  the curves in the
$\hsi-\mu'$ plane on which the equality $\hsi=\nnu T^{MB}({\bf 0},{\bf 0},{\bf
0};0)$ holds, as well as two curves on which the chemical potential $\mu$ is
constant, both at a temperature of $30 \mu K$. The straight line $\mu'=-\hsi$
corresponds to the conditions where the Hugenholtz-Pines theorem is satisfied.
The area $\hsi>\mid\mu'\mid$ is of course not part of the solution space since
it would imply a complex (unstable) dispersion relation. By calculating the
thermodynamic potential at the intersections of the lines we infer the relation
between $\Omega$ and the condensate density $\nnu$ for a given chemical
potential.

This is shown in Fig.\ 3 for the same two values of the chemical potential.
{}From this plot we find that the $\nnu$ minimizing the thermodynamic potential
for given $\mu$ indeed corresponds to the situation where the Hugenholtz-Pines
theorem is satisfied, i.e.\ $\mu'=-\hsi$. Note that this is a result of the
minimization procedure and is not assumed beforehand. The Hugenholtz-Pines
relation however turns out not to hold in every situation. For extreme
conditions, very close to zero temperature, the position of the minimum is
shifted slightly from the Hugenholtz-Pines line. (See also Fig.\ 13 in the next
subsection.) At $T=0$ one can, starting from Eq.\ (28) with general $\hsi$ and
$\mu'$, prove also analytically that this is the case. The deviation from the
Hugenholtz-Pines condition is due to an overestimate of the effective
interaction between the atoms. Indeed, a smaller value of the $T-$matrix,
together with a larger value of the condensate density to ensure that the
chemical potential $\mu$ remains the same, shifts the minimum towards the
Hugenholtz-Pines line. This overestimate of the effective interaction is due to
the neglect of the influence of the coherence factors in the expression for the
many-body $T-$matrix in Eq.\ (19), i.e.\ due to putting $u_{\bf k}=1$ and
$v_{\bf k}=0$. Taking these coherence factors into account will cause a
decrease of the $T-$matrix which presumably solves the problem that the minimum
of the $\Omega-\nnu-$curve does not coincide with the Hugenholtz-Pines
criterion. The region where the neglect of the coherence factors becomes
important corresponds to the condition $na\Lambda^2\stackrel{>}{\sim} 1$, which
is indeed in the extremely low temperature range where our numerical results
show a minimum in the thermodynamic potential that is slightly shifted from the
Hugenholtz-Pines line. Since this temperature range will be very difficult to
obtain experimentally, we will not consider it further here.

Outside this range, all the properties of the system can now be calculated
accurately. In Fig.\ 4 we plot the condensate fraction as a function of
temperature for a density of $10^{14}$ $cm^{-3}$. The inset shows the behavior
near the critical temperature following from our calculation, as well as from a
calculation in which we have replaced the many-body $T-$matrix by the two-body
$T-$matrix. The latter is essentially what has been done in earlier approaches
to the dilute Bose gas.\cite{STRAL} Using the two-body $T-$matrix however
produces a double-valuedness in this curve, leading inevitably to the
conclusion that there is a first-order phase transition. This double-valuedness
also shows up in the equation of state depicted in Fig.\ 5. Again, the equation
of state is single-valued if the many-body $T-$matrix is used. Our variational
approach has therefore succesfully resolved the long-standing problem
concerning the double-valuedness in the equation of state. It shows that it is
essential to include the effect of the surrounding medium on the collision of
two hydrogen atoms, which leads effectively to a temperature dependent
scattering amplitude and a softening of the dispersion near the critical
temperature. We calculated the effect of the medium on the effective scattering
length explicitly for the above situation where $n=10^{14}$ $cm^{-3}$, and show
in Fig.\ 6 the ratio of the many-body $T-$matrix to the two-body $T-$matrix as
a function of temperature. The fact that the many-body $T-$matrix, and
therefore the effective interaction of the atoms, goes to zero at the critical
temperature is correct when neglecting momentum and frequency dependencies as
can be shown from an exact renormalization-group equation for the four-point
vertex function.\cite{ZINNJ}

However, the many-body $T-$matrix $T^{MB}({\bf k},{\bf k'},{\bf K};z)$ does not
go to zero for nonzero momenta and frequencies. This explains the feature seen
in the inset of Fig.\ 7, where we plot the $p-n^{-1}$ diagram resulting from
our calculations. Near the critical density the curve shows a van der Waals
loop, which, by means of a Maxwell construction, would still imply a
first-order phase transition. This behavior is due to the fact that the
thermodynamic potential Eq.\ (28), and hence the pressure $p=-\Omega/V$, always
contains the many-body $T-$matrix evaluated at zero momenta and energies.
Therefore, the van der Waals loop shown in Fig.\ 7 is solely a consequence of
the behavior of the ratio $T^{MB}({\bf 0},{\bf 0},{\bf 0};0)/T^{2B}({\bf
0},{\bf 0};0)$ near the critical temperature (see Fig.\ 6). It would not have
been present had we used a more accurate expression for the thermodynamic
potential, which in principle contains integrals over the many-body $T-$matrix
evaluated at finite momenta and energies. Hence, the first-order phase
transition implicate from Fig.\ 7 is an artefact of our long wavelength
approximation which breaks down near the critical temperature. We stress
however that the softening of the dispersion discussed earlier, resolving the
problems related to the double-valuedness in the equation of state near the
critical temperature in contrast is {\em not} an artefact because the
dispersion relation {\em does} contain $T^{MB}({\bf 0},{\bf 0},{\bf 0};0)$ in
the long wavelength limit.

Finally in Fig.\ 8 we show the specific heat $C_V$ as a function of
temperature, again for a density of $10^{14}$ $cm^{-3}$. Close to $T_c$ our
results in principle show an artificial discontinuity which is due to the
first-order phase transition we find as explained above, and is therefore not
explicitly shown. This concludes our discussion of the three-dimensional
system, and we now turn to the two-dimensional case.

\subsection{The two dimensional case}
In Fig.\ 9 we show the solution curves of $\hsi=\nnu T^{MB}({\bf 0},{\bf
0},{\bf 0};0)$ for three different values of $\nnu a^2$ at a temperature of 50
$mK$, as well as two curves on which the chemical potential $\mu$ is constant.
In addition to the area $\hsi>\mid\mu'\mid$, also the Hugenholtz-Pines line
itself is not part of the solution space since it would imply a gapless
dispersion and therefore infrared divergences. This is also the cause of the
essential difference of Figs.\ 9 and 2. To avoid infrared divergences in the
first case, the solution curves of $\hsi=\nnu T^{MB}({\bf 0},{\bf 0},{\bf
0};0)$ are forced to start in the origin of the $\mu'-\hsi$ plane, in contrast
to what happens in three dimensions (cf. Fig.\ 2). This is a crucial point of
difference and results in the fact that the curves of the thermodynamic
potential versus chemical potential all start in one and the same point,
corresponding to $\mu=\hsi=\mu'=0$, as shown in Fig.\ 10, where $\nnu$
increases going from curve 1 to curve 4.

{}From this graph the following is then immediately clear: For $\mu<0$ the
value of $\nnu$ yielding the lowest thermodynamic potential is zero, whereas
for $\mu>0$ the value of $\nnu$ has to be infinite to give the lowest value of
$\Omega$. Thus for $\mu>0$ we cannot fix the number of particles in the system
to a finite value, and we conclude that the variational approach signals the
onset of an instability at $\mu=0$, but cannot describe the phase below the
critical temperature. Above $T_c$ it describes the two dimensional Bose gas in
the ladder approximation. What we {\em can} do is to deduce from the condition
$\mu=0$ the relation between the critical temperature and density as it follows
from our variational approach. This is shown in Fig.\ 11, where we plot $n_c$
versus $T_c$. As expected we find that the critical temperature is
overestimated, and $n_c\Lambda_c^2$ is smaller than 4. This is consistent with
the fact that the variational approach is of a mean-field type, since no
fluctuations in the background field have been taken into account.
Fluctuations, and in particular phase fluctuations associated with vortices,
should reduce the real critical temperature and result in the universal jump of
the superfluid density. This is schematically shown in Fig.\ 12, where we plot
the behavior of the superfluid density versus temperature. Only for
temperatures larger than $T_c^{MF}$, which is itself larger than $T_c^{KT}$ at
which the Nelson-Kosterlitz criterion $n_s\Lambda^2=4$ is satisfied, can we
apply our theory.

We would like to point out that $T_c^{MF}$ cannot be found by a straightforward
application of perturbation theory around the ideal Bose gas. Indeed,
calculating the free energy $F$ and the (normal) selfenergy in the ladder
approximation would result in
\begin{eqnarray}
\mu'=\mu-\hbar\Sigma_{11}=\mu-2nT^{MB}({\bf 0},{\bf 0},{\bf 0};0)
\end{eqnarray}
and in the equation of state
\begin{eqnarray}
n=\sum_{\bf k}N(\epsilon_{\bf k}-\mu')~,
\end{eqnarray}
where the latter is now obtained by a minimization of the free energy. At a
fixed temperature and density, we first of all determine $\mu'$ from Eq. (35)
and subsequently $\mu$ from Eq. (34). Clearly, this calculation can be
performed at any temperature and we just obtain the ideal gas result that a
phase transition occurs only at zero temperature. Notice however that for
temperatures larger than $T_c^{MF}$ the above procedure of minimizing the free
energy $F$ leads, as expected, to exactly the same results as our minimization
of the thermodynamic potential.

Finally we show in Fig.\ 13 the thermodynamic potential as a function of
condensate density for a fixed chemical potential in the case of zero
temperature where a real condensate can exist in the two-dimensional system. In
this case the solution curves of $\hsi=\nnu T^{MB}({\bf 0},{\bf 0},{\bf 0};0)$
do not start out in the origin of the $\mu'-\hsi-$plane, but essentially behave
as depicted in Fig.\ 2, because without thermal fluctuations there are no
infrared divergences  when $\mu'=-\hsi$. Again in this extreme situation we
find that the Hugenholtz-Pines relation is not satisfied, which, as in three
dimensions, is due to the neglect of the coherence factors $u_{\bf k}$ and
$v_{\bf k}$ in the equation for the many-body $T-$matrix by our variational
approach.

\section{Conclusions}
We performed a variational calculation for a weakly-interacting Bose gas and
applied this method to spin-polarized atomic hydrogen in both the
three-dimensional and the two-dimensional case. We showed that the variational
approach does not automatically lead to a consistent description of the system
under study, and that the number of diagrams that is being summed can be rather
restricted. In particular, the normal selfenergy is obtained in one-loop
approximation only. The anomalous selfenergy contains the ladder diagrams. We
consequently improved all relevant expressions to the $T-$matrix level.

In three dimensions this approach resolves the problem of double-valuedness in
the equation of state, that leads to the conclusion of a first-order phase
transition and is found in all previous approaches to the dilute Bose gas. The
essential ingredient to resolve this problem is the use of the many-body
$T-$matrix, which takes into account the fact that two particles collide in a
medium, and not in a vacuum. The (small) van der Waals loop we still find in
the $p-n^{-1}$ diagram is a result of the neglect of the momentum and energy
dependence of the $T-$matrix. The simplification in the coherence factors
$u_{\bf k}$ and $v_{\bf k}$ leads to an overestimate of the effective
interaction, and explains the fact that in extreme conditions, very far from
the critical temperature, the Hugenholtz-Pines relation is not satisfied. Using
the full many-body $T-$matrix $T^{MB}({\bf k},{\bf k'},{\bf K};z)$ will resolve
these issues. Not too close to the critical temperature or to zero temperature
the theory gives an excellent description of the dilute Bose gas. We also
calculated condensate fraction and specific heat as a function of temperature.

In two dimensions the variational approach, having a mean-field nature, signals
only the onset of an instability in the theory, but unfortunately cannot
describe the phase below the critical temperature. This is connected to the
fact that within this variational approach it is not possible to find solutions
where the dispersion relation of the quasiparticles displays a gap, playing the
same role as the infrared cutoff in the theory with a quasi-condensate and
enabling us to describe the system also below the critical temperature. The
critical temperature we can infer from the occurence of the instability is an
overestimate of the true critical temperature due to the neglect of the crucial
phase fluctuations. In this respect it is important to note that introducing a
quasi-condensate $n_0(k_0)$, followed by a diagrammatic calculation of the
equation of state,\cite{USKT} does give correct information about the
Kosterlitz-Thouless phase and thus incorporates the essential physics of the
superfluid state. This in contrast to the mean-field variational approach
presented here.

Finally we want to remark that solving the problem of the dilute Bose gas using
the full $T-$matrix Eq.\ (18) is a highly nontrivial task due to infrared
difficulties even in three dimensions. A renormalization-group study however
will solve these difficulties and more importantly will also correctly describe
the critical region. A study along these lines is underway.

\section*{Acknowledgments}
We would like to thank P.M. Stevenson for helpful discussions that introduced
us to the variational approach to quantum field theory. We would also like to
thank Ike Silvera for the hospitality enjoyed by one of us (M.B.) during his
stay at Harvard University and for interesting discussions concerning the
ongoing experiments aimed at the observation of the Kosterlitz-Thouless
transition.

\newpage
\section*{Figure Captions}
\noindent
Fig.\ 1 Diagrammatic representations.\\
(a) Diagrammatic representation of the $T-$matrix equation.\\
(b) The normal and anomalous selfenergies as found in the variational approach
together with the expressions for the exact Green's functions. Note that
iteration of the equation for $\Sigma_{12}$ generates the ladder diagrams. This
is not the case for $\Sigma_{11}$.\\
\\
Fig.\ 2 Lines of constant chemical potential ((4) and (5), the first having the
largest chemical potential), and solution curves of the equation $\hsi=\nnu
T^{MB}({\bf 0},{\bf 0},{\bf 0};0)$ in the $\mu'-\hsi$ plane for (1) $\nnu
a^3=4.8~10^{-12}$, (2) $\nnu a^3=5.0~10^{-12}$ and (3) $\nnu a^3=5.2~10^{-12}$.
Also shown is the line on which the Hugenholtz-Pines relation is satisfied.\\
\\
Fig.\ 3 Thermodynamic potential versus condensate density for the same two
values of the chemical potential as in Fig.\ 2 and at a temperature of $30 \mu
K$. The dots indicate the starting points on the Hugenholtz-Pines line of the
corresponding curves in the $\mu'-\hsi$ plane. The graph shows that at the
minimum the Hugenholtz-Pines condition is indeed satisfied.\\
\\
Fig.\ 4 The condensate fraction as a function of temperature for a total
density of $10^{14} cm^{-3}$. The inset shows the behavior of the condensate
fraction near the critical temperature revealing the double-valuedness (curve
(1)) if one uses the two-body $T-$matrix, which is resolved if one uses the
many-body $T-$matrix that includes the effect of the medium on the collision
(curve (2)).\\
\\
Fig.\ 5 The equation of state for small $\mu'$ at the critical temperature for
a system with $n=10^{14} cm^{-3}$, showing the double-valuedness discussed also
in Fig.\ 4.\\
\\
Fig.\ 6 The ratio of the many-body $T-$matrix to the two-body $T-$matrix at a
density of $10^{14} cm^{-3}$, showing that $T^{MB}({\bf 0},{\bf 0},{\bf 0};0)$
goes to zero at the critical temperature.\\
\\
Fig.\ 7 Pressure versus inverse density at $T=50 \mu K$. The inset shows a
small van der Waals loop present around the critical density (see text).\\
\\
Fig.\ 8 Heat capacity versus temperature when $n=10^{14} cm^{-3}$. Note that
$3nk_B/2$ corresponds to 0.002071 $JK^{-1}m^{-3}$.\\
\\
Fig.\ 9 Lines of constant chemical potential ((4) and (5), the first having the
largest chemical potential), and solution curves of the equation $\hsi=\nnu
T^{MB}({\bf 0},{\bf 0},{\bf 0};0)$ in the $\mu'-\hsi$ plane for (1) $\nnu
a^2=10^{-5}$, (2) $\nnu a^2=5~10^{-5}$ and (3) $\nnu a^2=10^{-4}$. Also
indicated is the Hugenholtz-Pines line.\\
\\
Fig.\ 10 Thermodynamic potential versus chemical potential at a temperature of
$50 mK$ and (1) $\nnu a^2=0$, (2) $\nnu a^2=10^{-5}$, (3) $\nnu a^2=10^{-4}$
and (4) $\nnu a^2=5~10^{-4}$. The dot indicates the starting point of the
corresponding curves in the $\mu'-\hsi$ plane, which lies in the origin for all
values of $\nnu a^2$.\\
\\
Fig.\ 11 Critical density versus temperature found from the criterion $\mu=0$,
the point at which the variational approach signals the occurrence of an
instability.\\
\\
Fig.\ 12 Schematic behavior of superfluid density versus temperature. The full
line results from an RG-analysis, that leads to the exact critical temperature
$T_c^{KT}$, obeying $n_s\Lambda_c^2=4$. The dashed curve corresponds to a
mean-field result with a critical temperature $T_c^{MF}$. Below $T_c^{MF}$ the
variational approach does not give information obout the properties of the
system.\\
\\
Fig.\ 13 Thermodynamic potential versus condensate density for fixed chemical
potential at zero temperature, showing that the Hugenholtz-Pines condition,
valid only at the endpoint of the curve, is not satisfied in the minimum of
this curve.\\

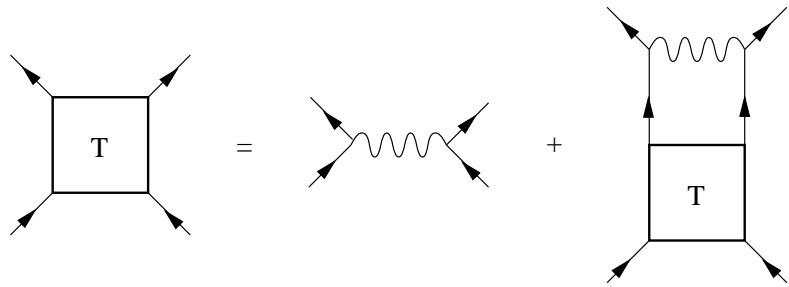
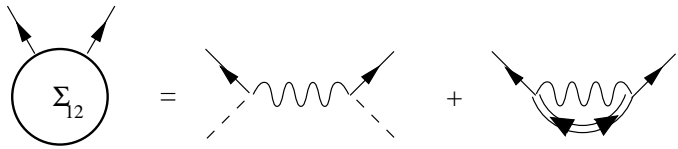
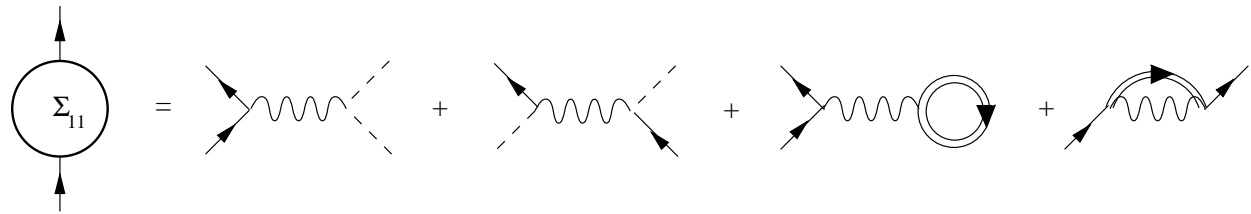
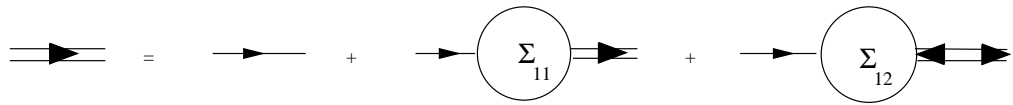
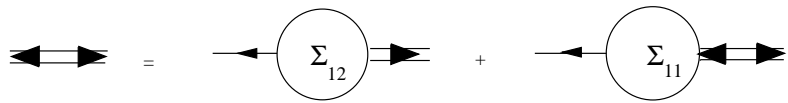
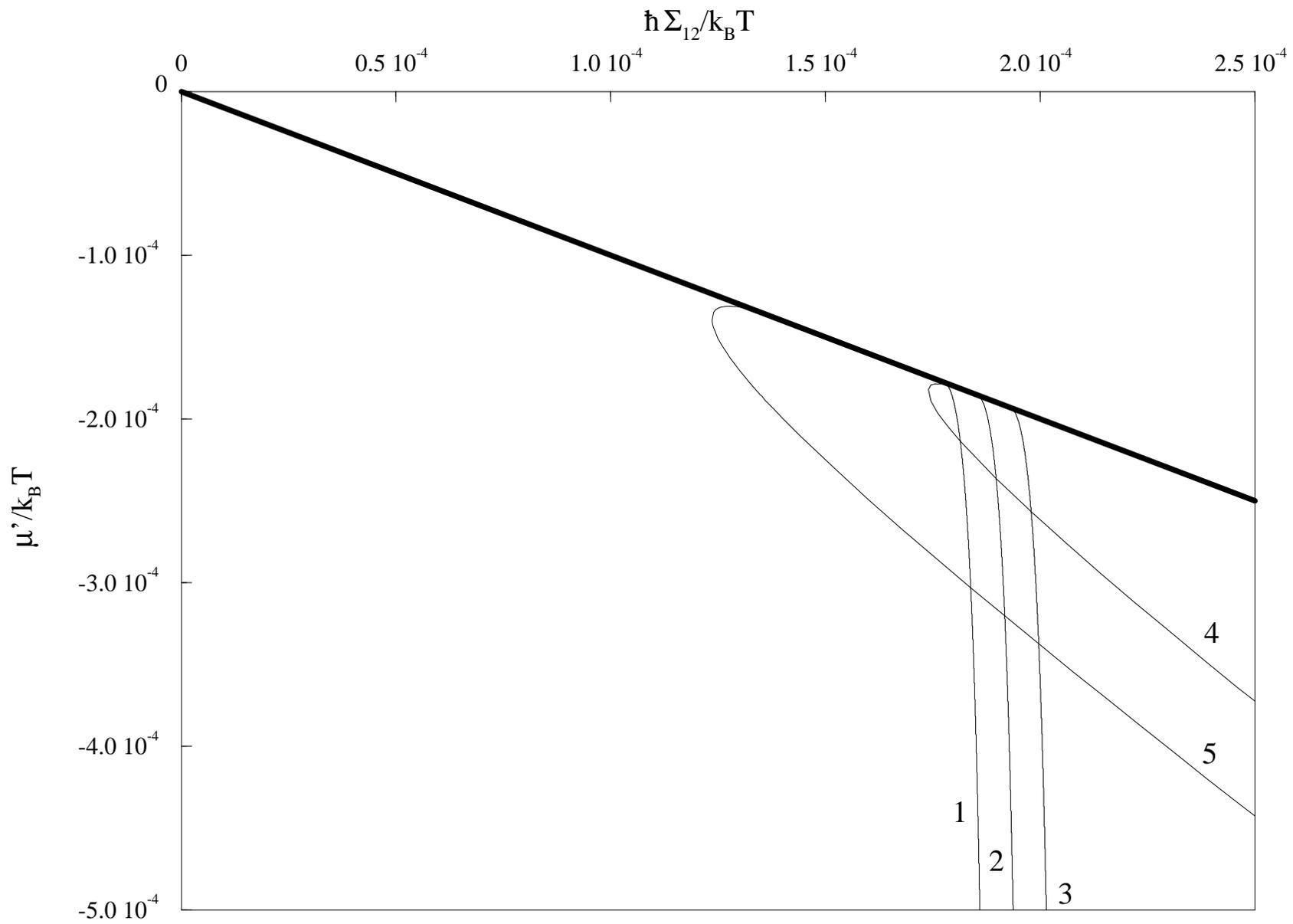
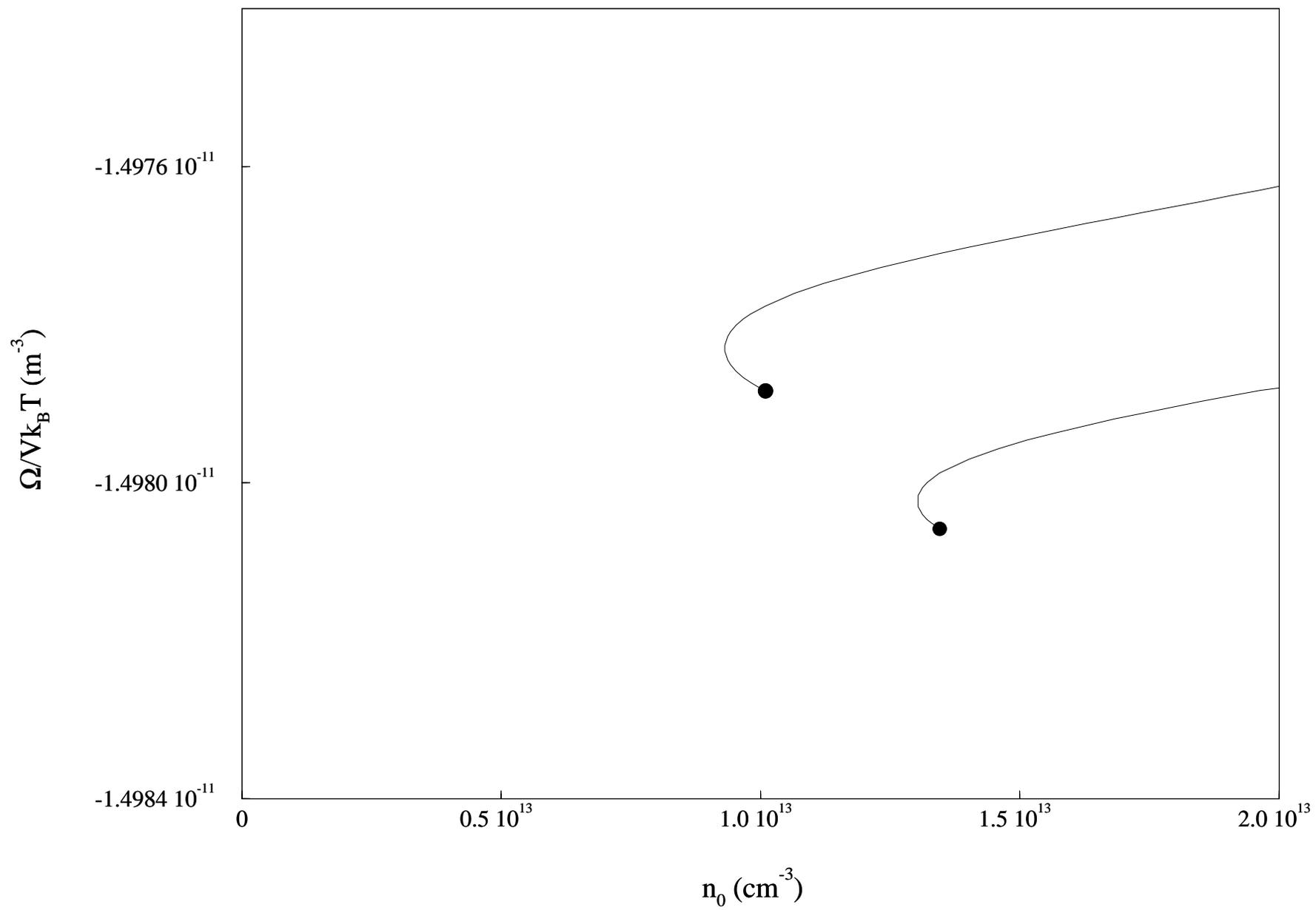
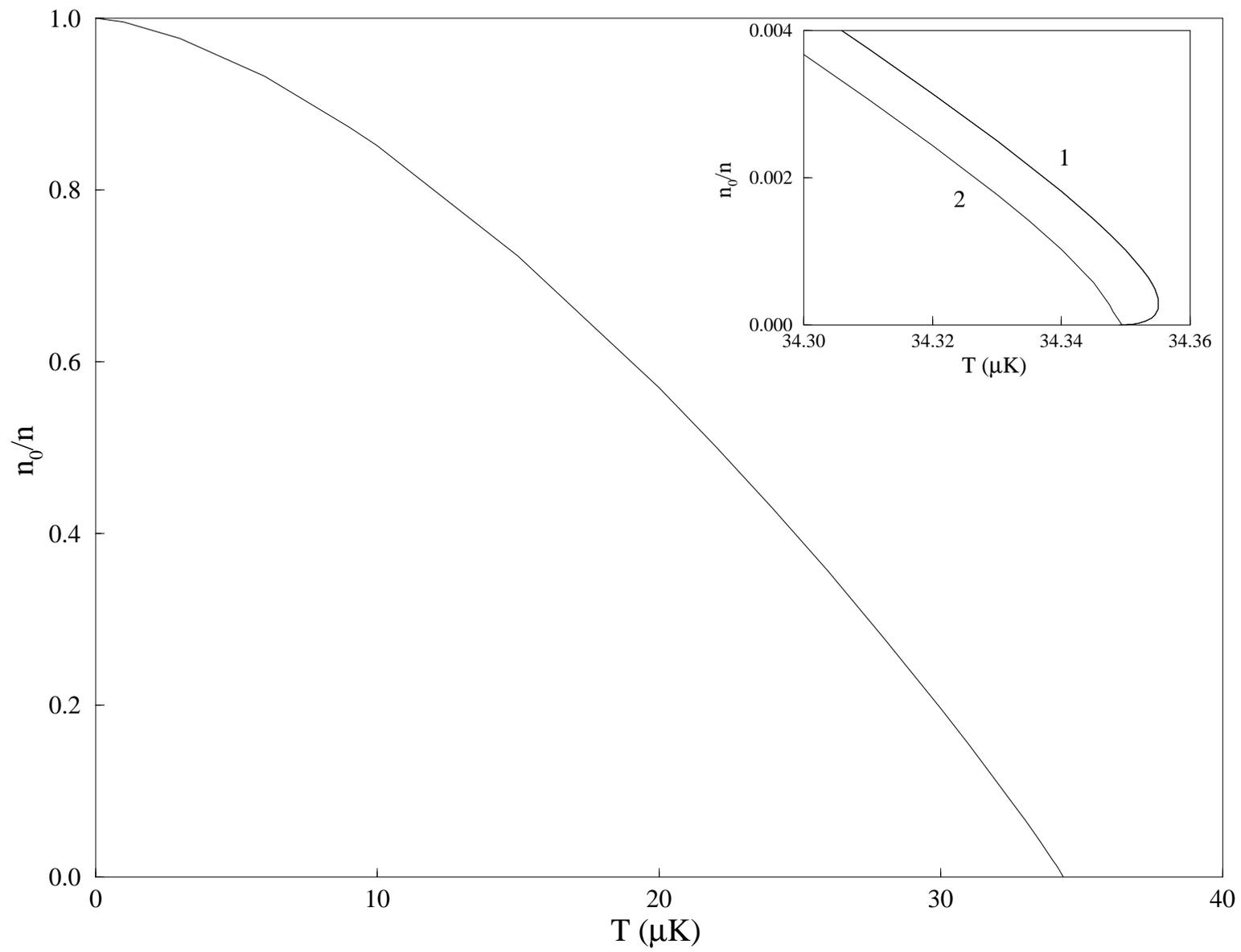
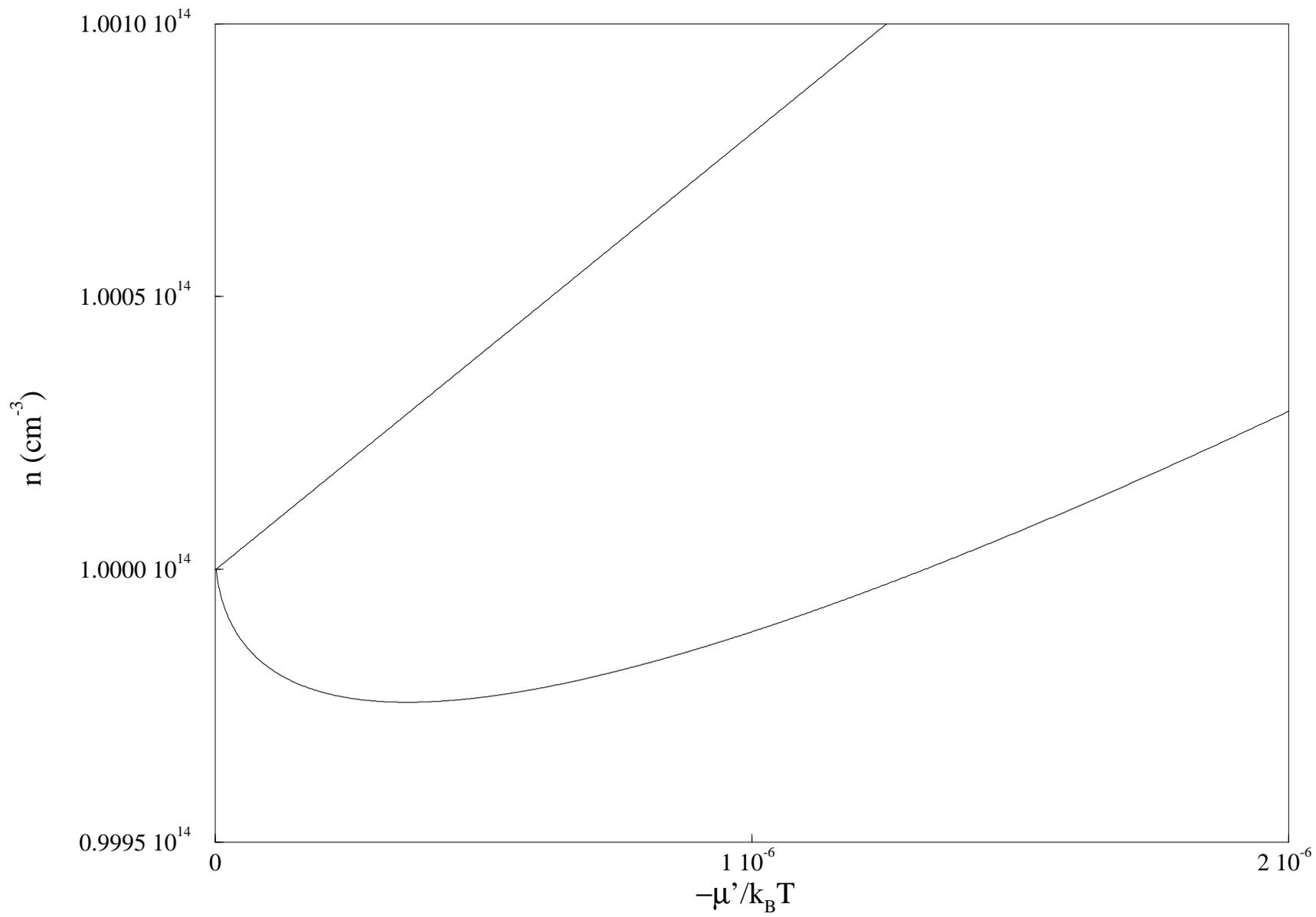
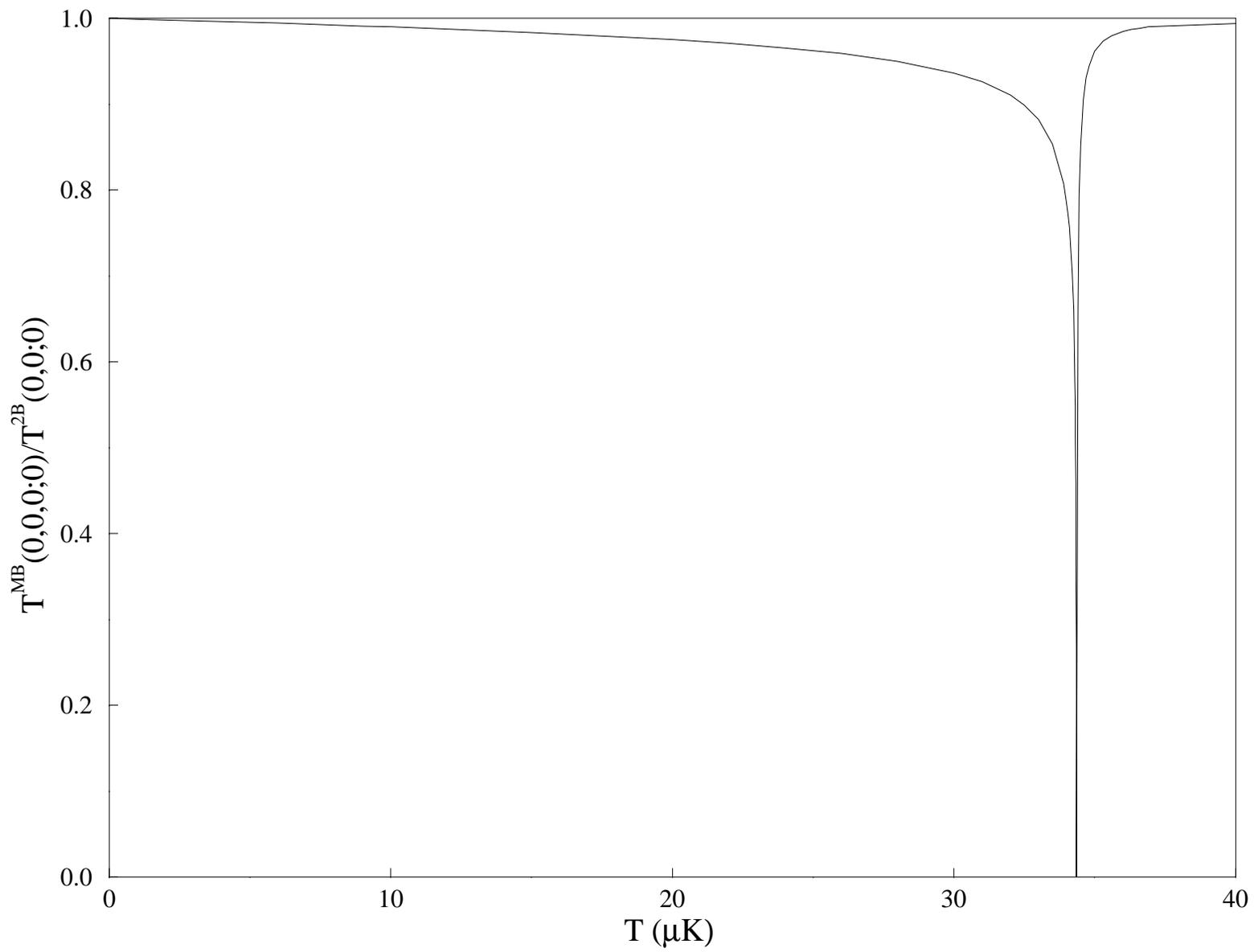
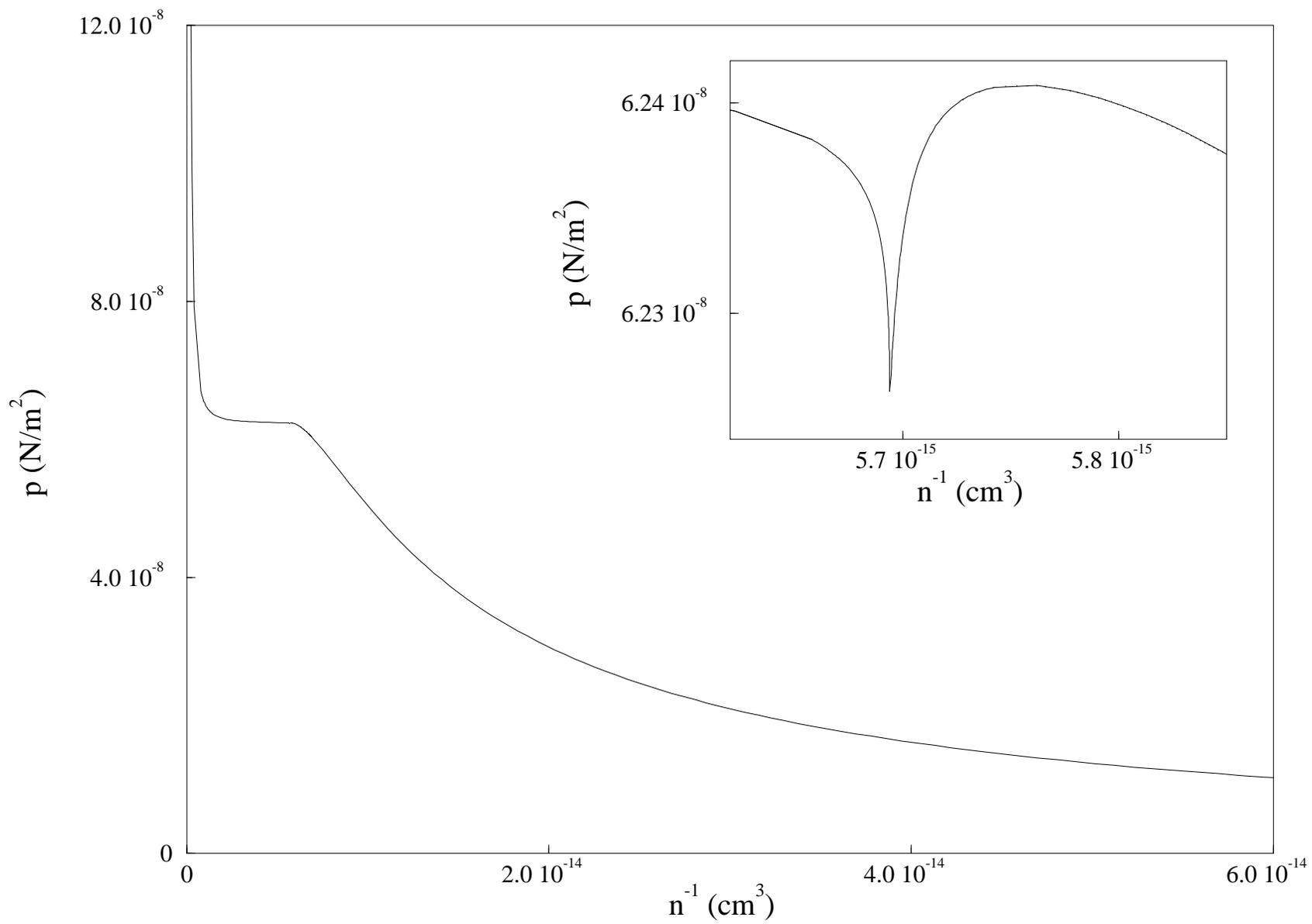
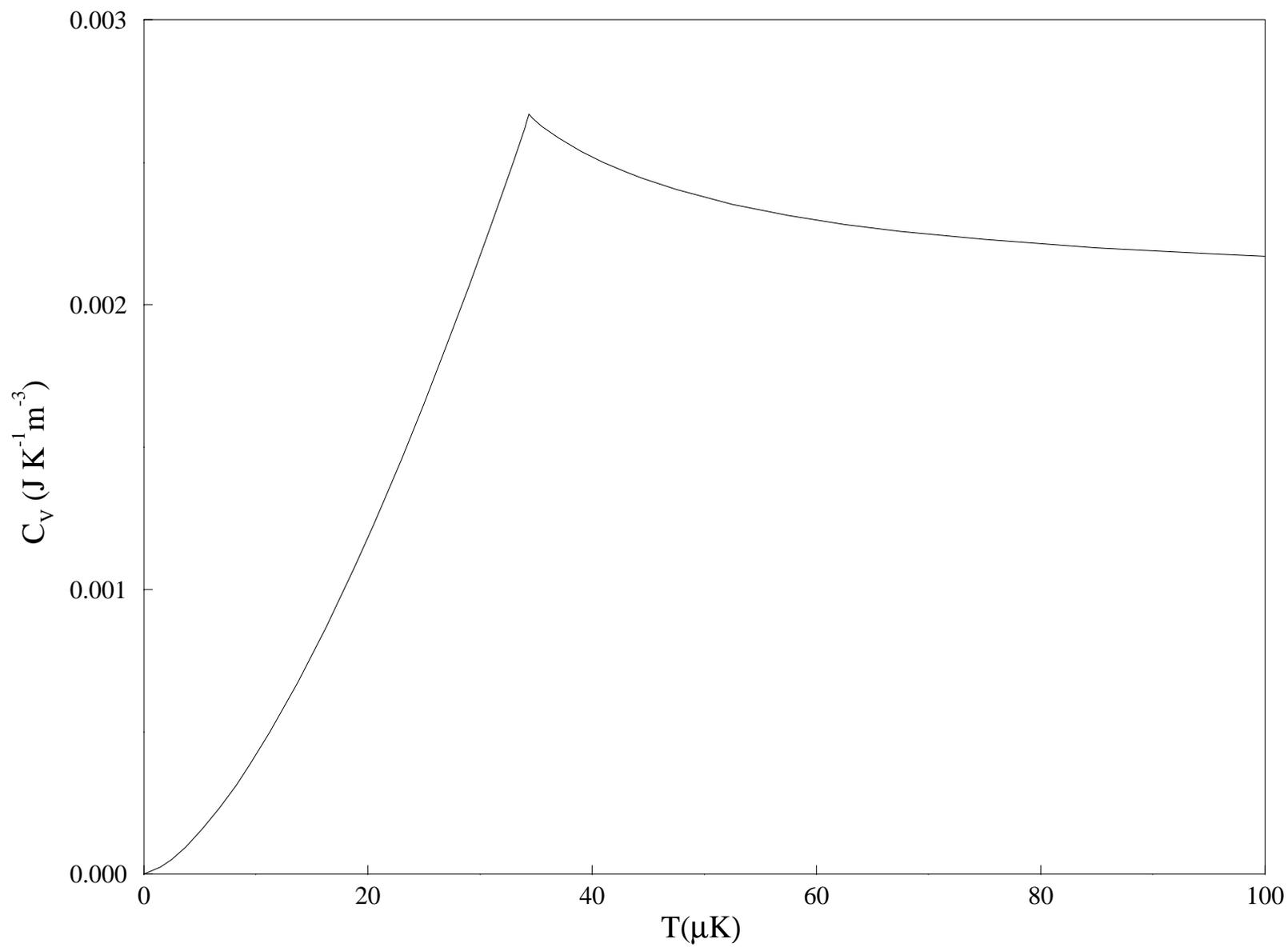
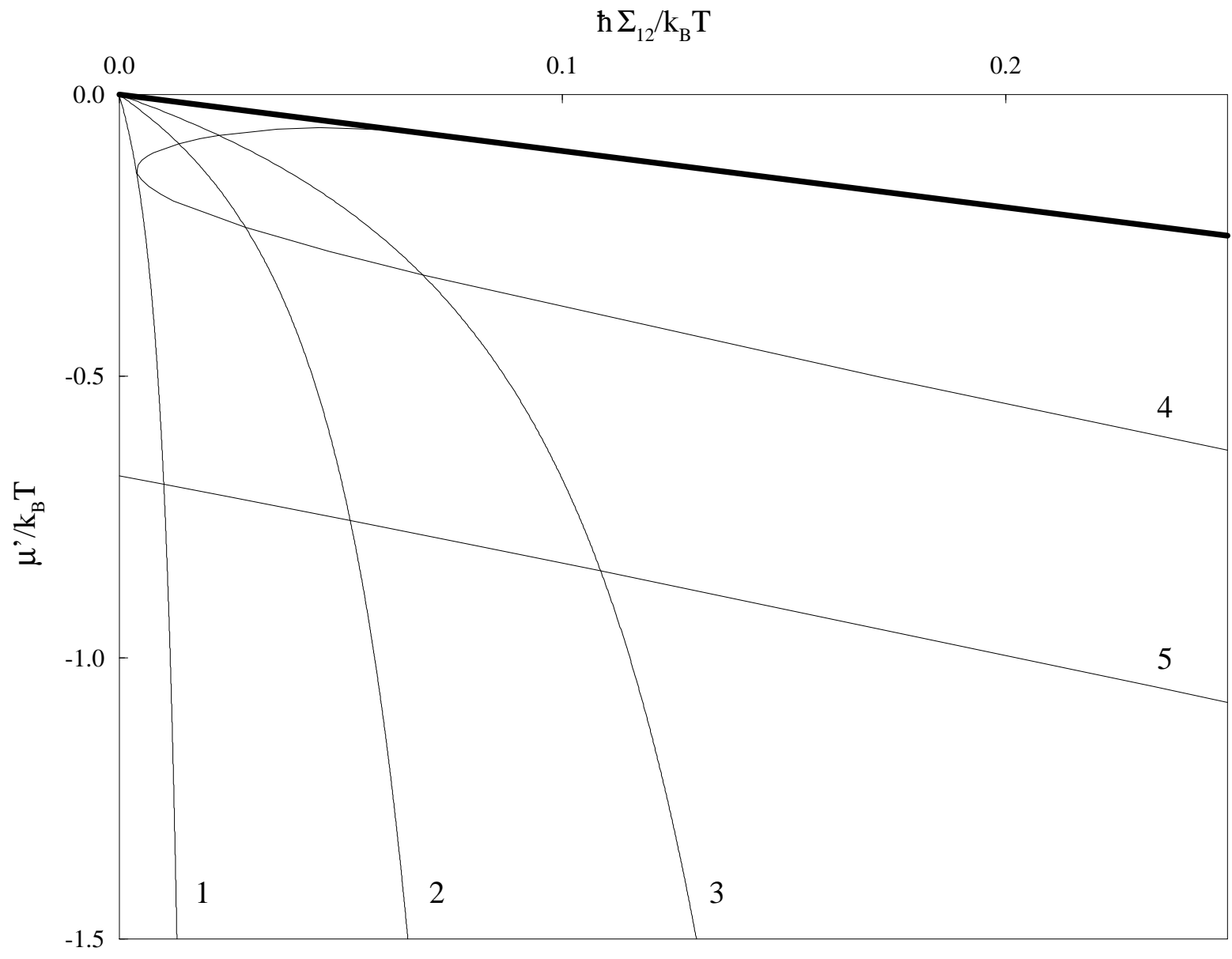
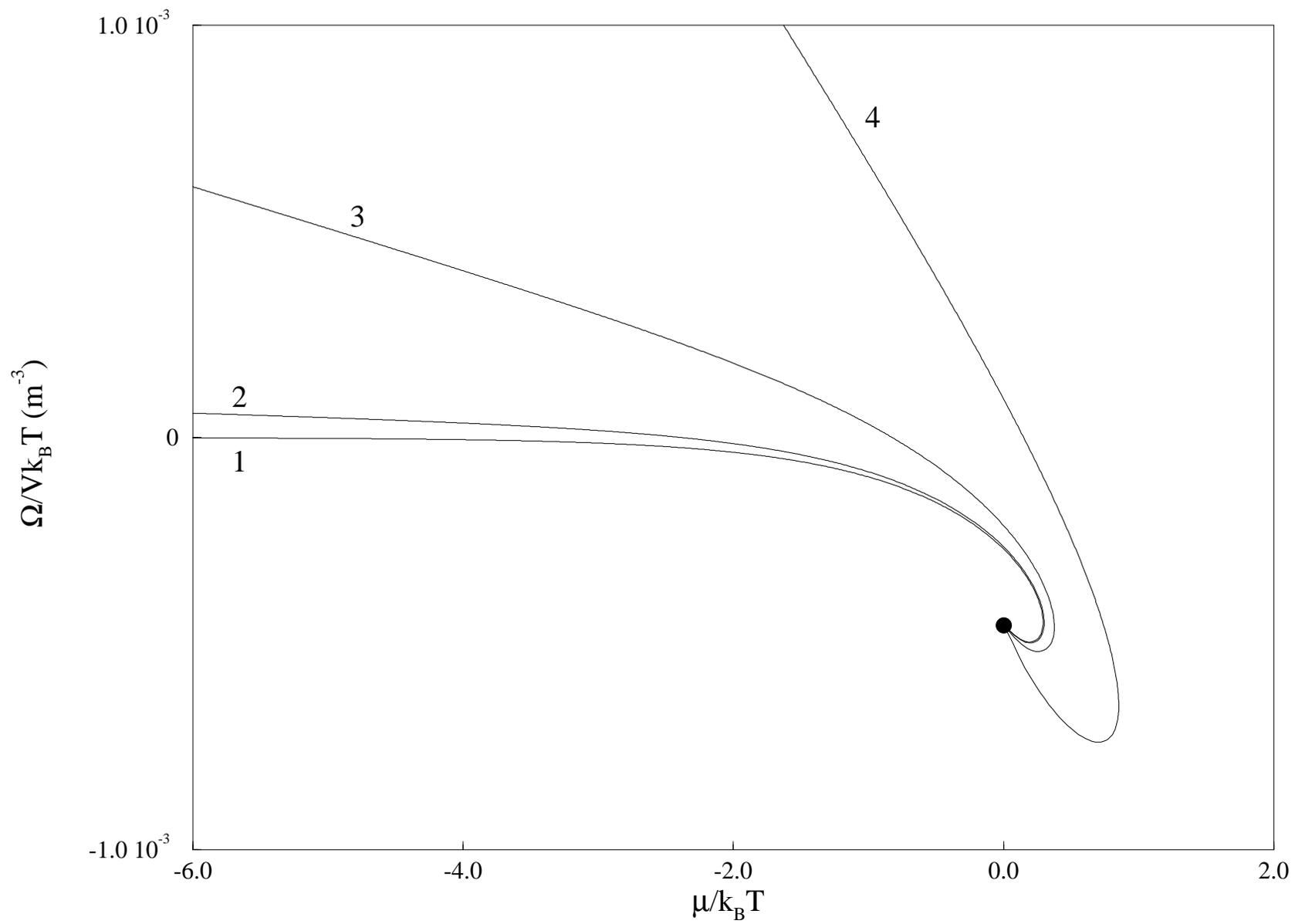
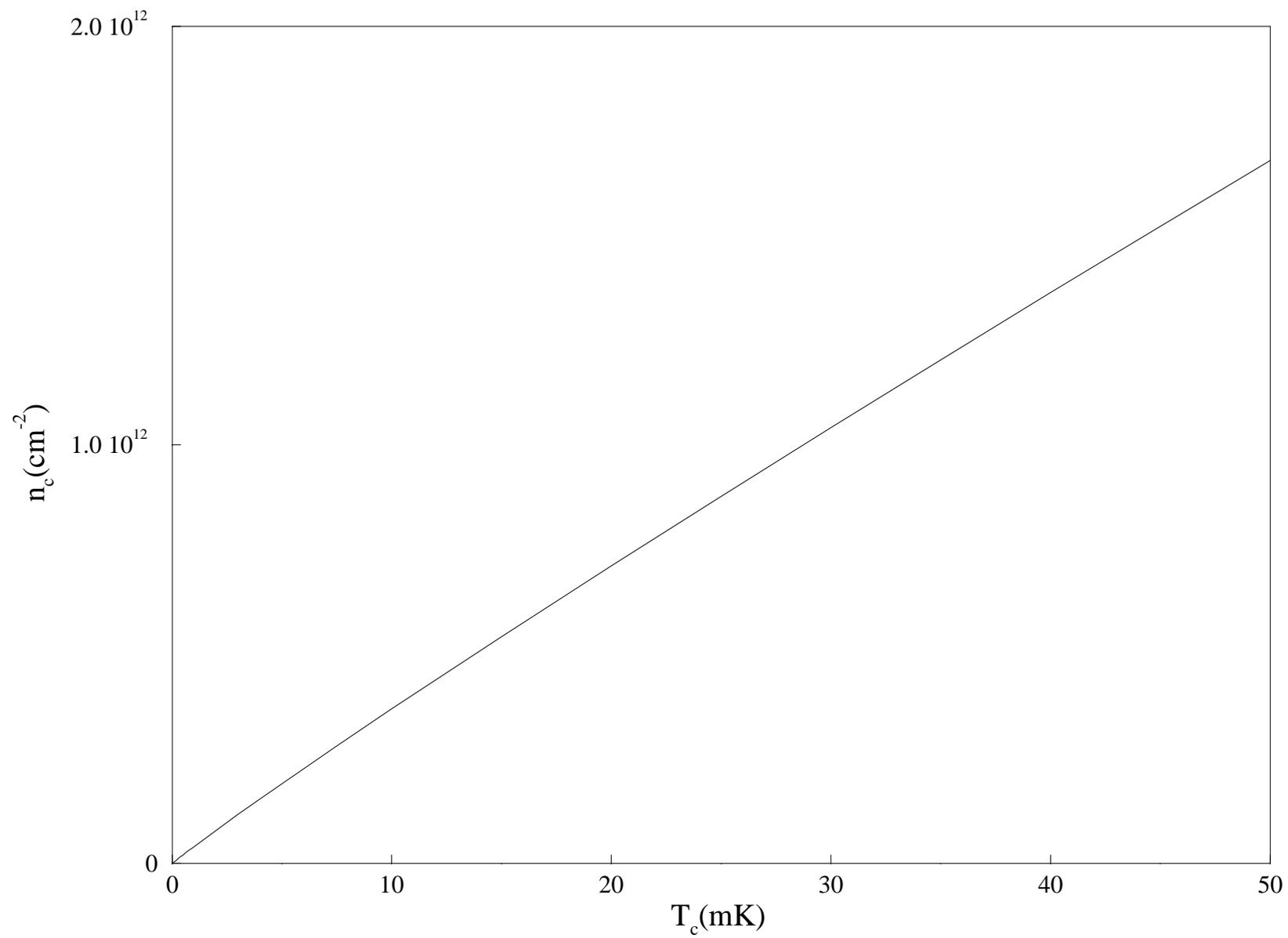
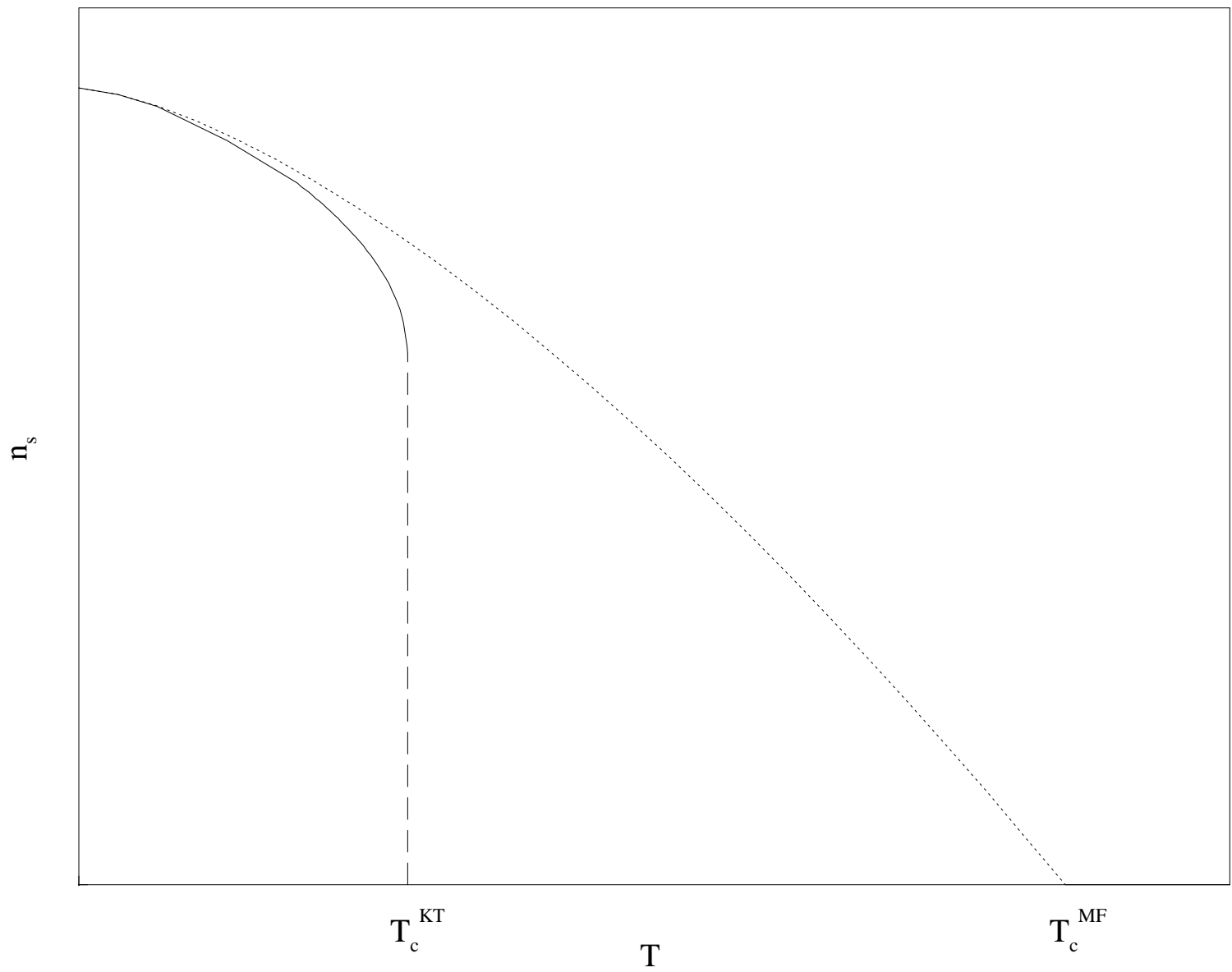
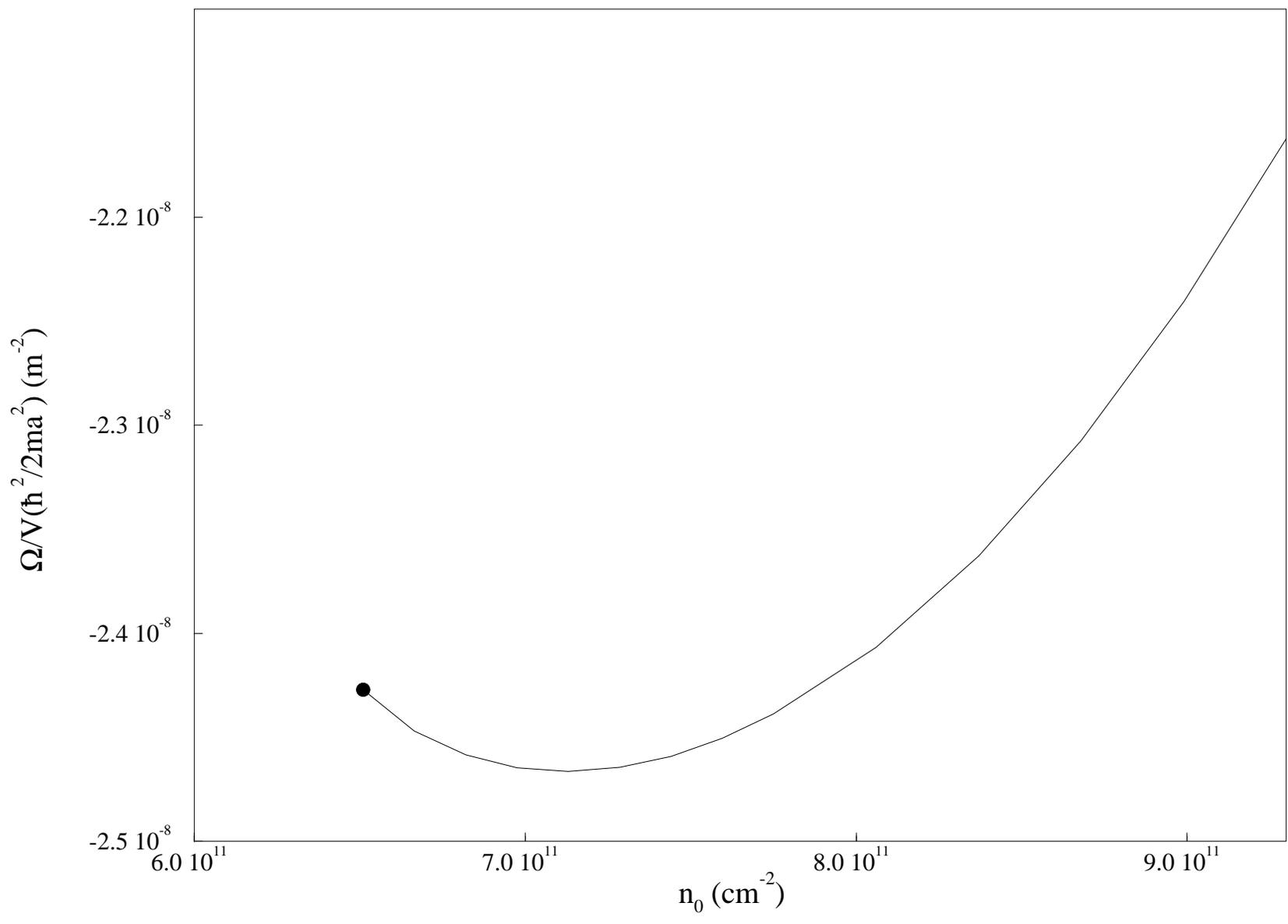
\end{document}